\documentclass{article}
\usepackage{iclr2027_conference,times}
\usepackage[T1]{fontenc}

\usepackage{amsmath,amsfonts,bm}

\def\eqref#1{equation~\ref{#1}}

\def\1{\bm{1}}

\DeclareMathAlphabet{\mathsfit}{\encodingdefault}{\sfdefault}{m}{sl}
\SetMathAlphabet{\mathsfit}{bold}{\encodingdefault}{\sfdefault}{bx}{n}

\usepackage{amsmath,amssymb,array,booktabs,graphicx,microtype,adjustbox,multirow}
\usepackage{siunitx}
\newcommand{\best}[1]{\multicolumn{1}{c}{\bfseries\boldmath #1}}
\newcommand{\second}[1]{\multicolumn{1}{c}{\underline{#1}}}
\usepackage{url,hyperref,enumitem,placeins}
\hypersetup{hidelinks}
\graphicspath{{shared/figures/}}

\makeatletter
\let\RNAoriginalmakecaption\@makecaption
\long\def\@makecaption#1#2{%
  \begingroup
  \setbox0=\vbox{\hsize=\linewidth\noindent #1: #2\par}%
  \typeout{RNA-CAPTION-AUDIT|#1|height=\the\ht0|depth=\the\dp0|baseline=\the\baselineskip}%
  \endgroup
  \RNAoriginalmakecaption{#1}{#2}%
}
\makeatother

\title{RNA Design via Conditioned Flow Matching and Finite-Policy Reinforcement Learning}
\author{\makebox[\dimexpr\textwidth-2\tabcolsep\relax][c]{Zefeng Lin$^{1}$\quad Xianyong Fang$^{3}$\quad Tianfan Fu$^{2}$\quad Xiaohua Xu$^{1}$}}
\begin{document}
\maketitle
\begin{center}\small 
$^{1}$School of Computer Science and Technology, University of Science and Technology of China, Hefei, Anhui, China\par
$^{2}$State Key Laboratory for Novel Software Technology at Nanjing University, School of Computer Science, Nanjing University, Nanjing, Jiangsu, China\par
$^{3}$School of Computer Science and Technology, Anhui University, Hefei, Anhui, China\par
\texttt{zflin@mail.ustc.edu.cn}; \texttt{fangxianyong@ahu.edu.cn}\par
\texttt{futianfan@nju.edu.cn}; \texttt{xiaohuaxu@ustc.edu.cn}
\end{center}

\begin{abstract}
RNA design aims to identify sequences that fold into specified secondary structures. Existing methods formulate the task as target-specific search or conditional generation. 
However, natural RNA evolution proceeds through sequence variation and selection, with compensatory substitutions, whereas these methods do not explicitly model this process. 
To address this limitation, we propose a two-stage framework comprising RNA Inverse-Folding Flow (\textbf{RNA-IFlow}) and \textbf{RNA-IFlow-RL}. RNA-IFlow uses structure-conditioned Dirichlet Flow Matching to model coordinated variation across the sequence, while RNA-IFlow-RL maps the learned flow to a pairing-preserving finite policy and refines it with thermodynamic feedback.
Our framework achieves leading performance on multiple benchmarks, reaching 85.19\% Pass@1 on Rfam-27. Further analyses reveal thermodynamic gains, policy dynamics, and robustness across settings.
Our work couples coordinated variation with thermodynamic selection, offering a novel paradigm for RNA design.
\end{abstract}

\section{Introduction}
\label{sec:introduction}
RNA is a programmable polymer of ribonucleotides that folds into hierarchical structures, where secondary structure captures the intramolecular base-pairing topology of the RNA chain~\citep{lorenz2011vienna,zadeh2011nupack}.
RNA design, also known as inverse folding, aims to identify nucleotide sequences that fold into a specified secondary structure and provides a computational route to design RNAs with desired structural properties for biotechnology and therapeutic applications~\citep{andronescu2004rnassd,zadeh2011nupack,ward2023fitness,li2025drag}.
Recent progress in AI-assisted mRNA therapeutics further highlights the broad potential of RNA design for applications such as personalized cancer vaccines and precision RNA therapeutics~\citep{fieldhouse2026moderna}.

\begin{figure}[!t]
\centering
\includegraphics[width=\textwidth]{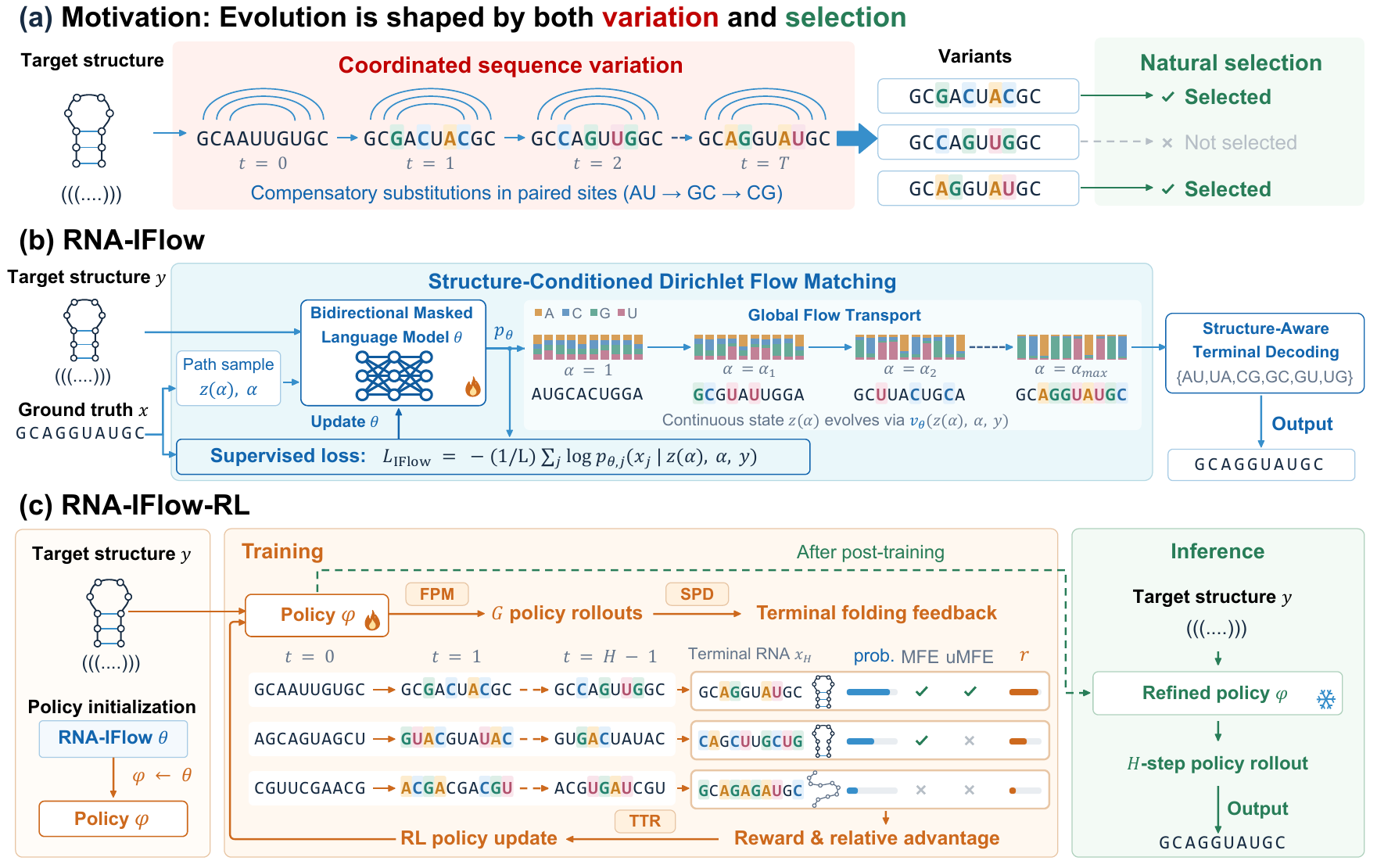}
\caption{
Overview of the two-stage framework. (a) illustrates the variation--selection motivation. (b) shows RNA-IFlow supervised training. (c) shows RNA-IFlow-RL training and inference. 
}
\label{fig:method_overview}
\end{figure}

Existing methods mainly solve RNA design through target-specific search, learned optimization, or conditional generation~\citep{li2025drag,zhou2026fastdesign,gautam2026rnadesignlm}.
Although these methods achieve substantial advances, they typically formulate RNA design as search policy optimization or structure-conditioned generation, which differs from the variation--selection process observed in natural RNA evolution.
In this process, sequence variation explores alternative nucleotide configurations, while compensatory substitutions at paired sites can preserve base pairing as mutations accumulate~\citep{chen1999compensatory,dutheil2010epistasis}. Selection then favors mutants that better satisfy structural and functional constraints, as illustrated in Figure~\ref{fig:method_overview}(a).
This observation suggests a natural modeling perspective for RNA design. We can first model structure-conditioned sequence variation, and then refine the resulting sequence distribution through thermodynamic selection.

However, modeling RNA design based on this perspective faces three fundamental challenges: 
(1) How can coordinated sequence variation be modeled over the complete RNA sequence rather than through independent or strictly sequential nucleotide decisions?
(2) How can a selection process be introduced so that generated variants are progressively biased toward thermodynamically favorable folds?
(3) How can compensatory substitutions at paired sites be preserved during variation, such that sequence changes remain consistent with the target secondary structure?

To address these challenges, we propose a two-stage framework comprising \textbf{RNA-IFlow} and \textbf{RNA-IFlow-RL}, which combines flow matching (FM) and reinforcement learning (RL) for RNA design (Figure~\ref{fig:method_overview}).
For the first challenge, RNA-IFlow applies the structure-conditioned Dirichlet FM on a bidirectional masked language model backbone to model coordinated variation over the complete RNA sequence state.
For the second challenge, RNA-IFlow-RL introduces RL to model thermodynamic selection over the learned flow predictor.
Specifically, it uses \textbf{Flow-to-Policy Mapping (FPM)} to convert the learned flow predictor into a finite policy with tractable transition probabilities, and \textbf{Thermodynamic Trajectory Refinement (TTR)} to propagate terminal folding feedback to the policy decisions that generate each sequence.
For the third challenge, we explicitly impose RNA structural constraints on both stages of our framework.
Specifically, RNA-IFlow applies structure-aware terminal decoding to assign valid states to target base pairs, while RNA-IFlow-RL extends this constraint throughout RL refinement by adopting \textbf{Structure-Preserving Policy Dynamics (SPD)}.

Our main contributions are summarized as follows:
\begin{enumerate}[leftmargin=*,nosep,label=(\arabic*)]
\item We propose a two-stage framework for RNA design that models the variation--selection process.
\item To address the modeling challenges, RNA-IFlow models coordinated sequence variation through structure-conditioned Dirichlet FM, while RNA-IFlow-RL converts the learned flow into a finite RL policy for thermodynamic selection. To integrate compensatory substitutions into the framework, we impose structural constraints on both stages.
\item Extensive experiments across multiple benchmarks demonstrate that our framework achieves leading performance, notably reaching 85.19\% Pass@1 on Rfam-27, compared with 81.48\% for RNA-Design-LM SL+RL. Further analyses show that our framework effectively models variation and selection, opening a promising direction for RNA design.
\end{enumerate} 
\section{Related Work}
\label{sec:related_work}

\paragraph{RNA Design.}
Existing RNA design methods mainly rely on target-specific search optimization, directly optimizing sequences according to structural or thermodynamic objectives. Representative approaches include NEMO, SAMFEO, SamplingDesign, and FastDesign~\citep{portela2018nemo,zhou2023samfeo,tang2026samplingdesign,zhou2026fastdesign}. Recently, learning-based and conditional generation methods have been introduced to generalize this optimization. LEARNA and DRAG use learned construction or mutation policies, while RNA-Design-LM and GoForth generate sequences conditioned on target structures~\citep{runge2019learna,li2025drag,gautam2026rnadesignlm,lindsey2026goforth}. 
However, these methods do not explicitly exploit the evolutionary information in RNA datasets.

\paragraph{Flow Matching for Biomolecular Generation.}
Flow matching (FM) provides a general framework for learning probability transport between distributions~\citep{lipman2023flow}. Dirichlet and discrete FM variants further extend this idea to categorical sequence spaces, realizing flow-based generation over discrete spaces~\citep{stark2024dirichlet,gat2024dfm}. Recently, flow-based models have been explored for biomolecular generation, including RNA sequence--structure co-design~\citep{nori2024rnaflow,ma2025riboflow}. 
These methods demonstrate the potential of FM methods for RNA design.

\paragraph{Reinforcement Learning for RNA Design.}
Reinforcement learning (RL) has been explored for optimizing RNA sequences toward structural and thermodynamic objectives. 
LEARNA and DRAG use RL for RNA sequential construction and mutation-based optimization, while RNA-Design-LM applies RL to optimize a trained conditional generator~\citep{runge2019learna,li2025drag,gautam2026rnadesignlm}. 
Building on these advances, we adopt RL to model thermodynamic selection.

\section{Method}
\label{sec:method}

\subsection{Problem Formulation} 
An RNA sequence is an ordered chain of nucleotides $x=(x_1,\ldots,x_L)\in\mathcal A^L$, where $\mathcal A=\{A,C,G,U\}$ denotes the four nucleotide types. Its target secondary structure $y$ describes the expected nucleotide base-pairing pattern in dot-bracket notation, with base-pair set $\mathcal P(y)$.
A thermodynamic model assigns free energy $\mathcal G(s;x)$ to each compatible structure $s\in\mathcal S(x)$, where $\mathcal S(x)$ denotes the set of valid secondary structures for sequence $x$.
The minimum-free-energy (MFE) structure set of $x$ is $\mathcal M(x)=\arg\min_{s\in\mathcal S(x)}\mathcal G(s;x)$, and the target probability is
\begin{equation}
p(y\mid x)=\frac{\exp[-\mathcal G(y;x)/(RT)]}
{\sum_{s\in\mathcal S(x)}\exp[-\mathcal G(s;x)/(RT)]},
\label{eq:target_probability}
\end{equation}
where $R$ is the molar gas constant and $T$ the absolute temperature; $p(y\mid x)=0$ when $y\notin\mathcal S(x)$. MFE success means $y\in\mathcal M(x)$, while unique-MFE (uMFE) success requires $\mathcal M(x)=\{y\}$. 
A successful design should not only make the target the unique MFE structure, but also favor it thermodynamically over competing conformations. We therefore formulate the objective of RNA design as
\begin{equation}
\max_{x\in\mathcal A^L}p(y\mid x)
\qquad\mathrm{s.t.}\qquad\mathcal M(x)=\{y\}.
\label{eq:design_objective}
\end{equation}
The goal of our framework is to learn the distribution of RNA sequences conditioned on a given target structure to obtain a successful design.
It proceeds in two stages. RNA-IFlow learns a structure-conditioned sequence distribution from supervised sequence--structure pairs. RNA-IFlow-RL then uses terminal thermodynamic selection to reshape this distribution toward successful designs. 

\subsection{RNA-IFlow}
As illustrated in Figure~\ref{fig:method_overview}(b), we present
\textbf{RNA-IFlow} as a structure-conditioned global flow in RNA sequence space to model coordinated sequence variation under structural constraints.
It combines \textbf{Structure-Conditioned Dirichlet Flow Matching}, \textbf{Global Transport}, and \textbf{Structure-Aware Terminal Decoding}. 
The three components respectively aim to learn structure-conditioned variation, enable coordinated sequence-wide generation, and ensure structurally valid decoding.

\paragraph{Structure-Conditioned Dirichlet Flow Matching.}
We first construct a continuous flow that captures structure-conditioned nucleotide variation.
Represent nucleotide $x_j$ by its one-hot vector $e(x_j)$ on the four-category simplex $\Delta^3=\{z\in\mathbb R_{\geq0}^4:\sum_{a\in\mathcal A} z_a=1\}$. Following Dirichlet Flow Matching~\citep{stark2024dirichlet}, we define the conditional Dirichlet probability path
\begin{equation}
q_\alpha(z_j\mid x_j)
=\operatorname{Dir}\!\left(z_j;\mathbf1+(\alpha-1)e(x_j)\right),
\label{eq:dirichlet_path}
\end{equation}
where $\mathbf1\in\mathbb R^4$ is the all-ones vector and
$\alpha\geq1$ serves as the continuous flow parameter and controls
concentration toward the clean nucleotide. 
For the complete sequence, we use the factorized path
$q_\alpha(z\mid x)=\prod_{j=1}^{L}q_\alpha(z_j\mid x_j)$.
To infer the direction of sequence variation along this path, given the complete state $z=(z_1,\ldots,z_L)$, level $\alpha$, and structure $y$, a bidirectional masked language model backbone $f_\theta$ predicts clean-nucleotide probabilities $p_{\theta,j}(a\mid z,\alpha,y)=\operatorname{Softmax}(f_\theta(z,\alpha,y)_j)_a$ for $a\in\mathcal A$. 
These structure-conditioned posteriors are then converted into a continuous velocity field
\begin{equation}
v_{\theta,j}(z,\alpha,y)=\mathsf P_{\mathrm{tan}}\!\left[
\sum_{a\in\mathcal A}p_{\theta,j}(a\mid z,\alpha,y)
 c_\alpha(z_{j,a})\bigl(e(a)-z_j\bigr)\right],
\label{eq:posterior_velocity}
\end{equation}
where $c_\alpha$ is the analytic conditional-flow coefficient and $\mathsf P_{\mathrm{tan}}$ projects onto the simplex tangent space. Their implementation is given in Appendix~\ref{app:theory}. 

\paragraph{Global Transport in RNA-IFlow.}
Having defined the structure-conditioned vector field, we next use it to transport the entire RNA sequence state.
RNA-IFlow initializes each position from $z_j\sim\operatorname{Dir}(\mathbf1)$ and evolves the complete sequence state through
\begin{equation}
\frac{dz_j(\alpha)}{d\alpha}
=
v_{\theta,j}(z(\alpha),\alpha,y),
\qquad j=1,\ldots,L.
\label{eq:iflow_transport}
\end{equation}
Because $v_{\theta,j}$ is predicted from the full sequence state and the shared target structure, all nucleotide positions are updated together under bidirectional context. This global transport allows coordinated sequence-wide variation to emerge throughout generation. 

\paragraph{Structure-Aware Terminal Decoding.}
Global transport generates a continuous terminal state on the nucleotide simplex, which must finally be converted into a discrete RNA sequence while respecting the target pairing constraints. 
We therefore decode unpaired sites by single-position argmax and each target pair by joint argmax over $\mathcal A_{\mathrm{pair}}=\{AU,UA,CG,GC,GU,UG\}$.
This structure-aware decoding maps the continuous flow state into a discrete RNA sequence while ensuring that every target base pair is assigned a valid pairing state.

\subsection{RNA-IFlow-RL}
As shown in Figure~\ref{fig:method_overview}(c), we present \textbf{RNA-IFlow-RL} to further introduce a thermodynamic selection process into RNA-IFlow.
It combines three designs: \textbf{Flow-to-Policy Mapping (FPM)} provides tractable discrete transitions, \textbf{Structure-Preserving Policy Dynamics (SPD)} preserves target-pair legality throughout generation, and \textbf{Thermodynamic Trajectory Refinement (TTR)} propagates terminal feedback to the decisions producing the final sequence.

\paragraph{Flow-to-Policy Mapping (FPM).}
We first convert the continuous RNA-IFlow representation into an executable discrete policy with tractable transition probabilities.
To reuse the trained RNA-IFlow predictor, FPM maps the current sequence $x_t\in\mathcal A^L$ to the conditional mean of the path in Equation~\ref{eq:dirichlet_path}:
\begin{equation}
\mu_\alpha(a)=\frac{\mathbf1+(\alpha-1)e(a)}{\alpha+3}
=\mathbb E_{Z\sim q_\alpha(\cdot\mid a)}[Z],
\qquad z_{t,j}=\mu_{\alpha_t}(x_{t,j}).
\label{eq:fpm_mean_bridge}
\end{equation}
Here $t$ indexes policy steps, $\alpha_t$ denotes the Dirichlet path level
associated with step $t$, and $j$ indexes nucleotide positions, while
$Z\in\Delta^3$ denotes a random simplex vector.
This mapping is the exact conditional mean of the Dirichlet path. Initialized from RNA-IFlow, $f_\phi$ predicts the clean-nucleotide posterior $p_{\phi,j}(a\mid z_t,\alpha_t,y)$ from $(z_t,\alpha_t,y)$, using the same parameterization as $p_{\theta,j}$.

Given these predictions, we construct a finite discrete transition kernel over structural units. 
Let $\mathcal U(y)$ denote the disjoint structural units induced by $y$,
with each unit corresponding to either an unpaired position or a target base pair,
and let $x_{t,u}$ denote the state of unit $u$ at step $t$.
For $u\in\mathcal U(y)$, SPD defines a resampling distribution $h_{\phi,u}$, giving
\begin{equation}
\begin{aligned}
\pi_{\phi,u}(x_{t+1,u}\mid x_t,\alpha_t,y)
=(1-\rho_t)\delta_{x_{t,u}}(x_{t+1,u})+\rho_t h_{\phi,u}(x_{t+1,u}\mid z_t,\alpha_t,y).
\end{aligned}
\label{eq:fpm_transition}
\end{equation}
For an $H$-step policy horizon, $\rho_t=1/(H-t)$ for $t=0,\ldots,H-1$, and $\delta$ denotes a point mass at the current unit state. This normalized transition kernel gives the exact transition probability under the constructed finite policy, enabling tractable policy optimization.

\paragraph{Structure-Preserving Policy Dynamics (SPD).}
While FPM makes RNA-IFlow transitions tractable, token-wise updates can still violate target base pairs at intermediate steps. SPD therefore constrains the action space of the structural units defined above: unpaired sites have four states in $\mathcal A$, and paired sites $(i,j)\in\mathcal P(y)$ have six states in $\mathcal A_{\mathrm{pair}}$.
Let $\widetilde p_{\phi,j}$ denote the temperature-scaled clean-nucleotide posterior. Unpaired units use $\widetilde p_{\phi,j}$ directly, while paired units construct a joint policy over legal base-pair states:
\begin{equation}
h_{\phi,(i,j)}(a,b\mid z_t,\alpha_t,y)
=\frac{\widetilde p_{\phi,i}(a\mid z_t,\alpha_t,y)\,
\widetilde p_{\phi,j}(b\mid z_t,\alpha_t,y)}
{\sum_{(c,d)\in\mathcal A_{\mathrm{pair}}}
\widetilde p_{\phi,i}(c\mid z_t,\alpha_t,y)\,
\widetilde p_{\phi,j}(d\mid z_t,\alpha_t,y)},
\label{eq:spd_pair_policy}
\end{equation}
for $(a,b)\in\mathcal A_{\mathrm{pair}}$. 
By restricting paired actions to $\mathcal A_{\mathrm{pair}}$, SPD preserves pairing legality at every policy step and extends the structural constraint to the full finite-policy trajectory.

\paragraph{Thermodynamic Trajectory Refinement (TTR).}
FPM and SPD establish a tractable, structure-preserving policy, but policy refinement still requires translating sequence-level thermodynamic outcomes into learning signals for the generation trajectory. TTR addresses this by evaluating terminal sequences and propagating their relative thermodynamic quality back through the trajectories.
For each target $y$, we sample $G$ trajectories, denoted by
$\tau^{(g)}=(x_0^{(g)},\ldots,x_H^{(g)})$, $g=1,\ldots,G$.
Each terminal sequence $x_H^{(g)}$ receives the reward
\begin{equation}
r^{(g)}
=
\beta_p\,p(y\mid x_H^{(g)})
+\beta_{\mathrm{MFE}}\,\mathbb I_{\mathrm{MFE}}(x_H^{(g)},y)
+\beta_{\mathrm{uMFE}}\,\mathbb I_{\mathrm{uMFE}}(x_H^{(g)},y),
\label{eq:ttr_reward}
\end{equation}
where
$\mathbb I_{\mathrm{MFE}}(x,y)=\mathbf1[y\in\mathcal M(x)]$,
$\mathbb I_{\mathrm{uMFE}}(x,y)=\mathbf1[\mathcal M(x)=\{y\}]$,
and $\beta_p$, $\beta_{\mathrm{MFE}}$, and $\beta_{\mathrm{uMFE}}$ are nonnegative weights.
Then, we compute the group-normalized advantage
\begin{equation}
A^{(g)}
=
\frac{r^{(g)}-\bar r_y}
{\sigma_y+\epsilon_{\mathrm{adv}}},
\label{eq:ttr_advantage}
\end{equation}
where $\bar r_y$ and $\sigma_y$ denote the group mean and population standard deviation, and $\epsilon_{\mathrm{adv}}>0$ ensures numerical stability. The same terminal advantage is assigned to every transition in $\tau^{(g)}$, allowing terminal thermodynamic selection to refine policy decisions along the entire generation trajectory.

\subsection{Training and Inference}
\label{sec:training_inference}

\paragraph{Training.}
We train our framework in two stages. In the first stage, RNA-IFlow learns structure-conditioned sequence variation by predicting clean nucleotides along
the Dirichlet probability path. For a sequence--structure pair
$(x,y)\sim\mathcal D_{\mathrm{SL}}$ where $\mathcal D_{\mathrm{SL}}$ denotes the supervised training set, we sample a path level $\alpha$ and a
simplex state $z\sim q_\alpha(\cdot\mid x)$, and optimize
\begin{equation}
\mathcal L_{\mathrm{IFlow}}(\theta)
=
\mathbb E_{\substack{
(x,y)\sim\mathcal D_{\mathrm{SL}},\;
\alpha\sim p(\alpha),\\
z\sim q_\alpha(\cdot\mid x)
}}
\left[
-\frac{1}{L}
\sum_{j=1}^{L}
\log p_{\theta,j}(x_j\mid z,\alpha,y)
\right].
\label{eq:iflow_training}
\end{equation}
Here $p(\alpha)$ is the training distribution over path levels. This objective trains the clean-nucleotide predictor that defines the RNA-IFlow vector field in Equation~\ref{eq:posterior_velocity}. 

In the second stage, RNA-IFlow-RL initializes its policy parameters from the trained RNA-IFlow model with $\phi\leftarrow\theta$, and further optimizes the resulting finite policy using the terminal advantages $A^{(g)}$ defined by TTR. To evaluate an executed transition under the current and rollout policies, we first form its joint probability over all structural units:
\begin{equation}
\begin{aligned}
\pi_\phi(x_{t+1}^{(g)}\mid x_t^{(g)},\alpha_t,y)
&=
\prod_{u\in\mathcal U(y)}
\pi_{\phi,u}
(x_{t+1,u}^{(g)}\mid x_t^{(g)},\alpha_t,y),
\\
\omega_t^{(g)}(\phi)
&=
\frac{
\pi_\phi(x_{t+1}^{(g)}\mid x_t^{(g)},\alpha_t,y)
}{
\pi_{\phi_{\mathrm{old}}}
(x_{t+1}^{(g)}\mid x_t^{(g)},\alpha_t,y)
}.
\end{aligned}
\label{eq:ttr_ratio}
\end{equation}
Here $\phi_{\mathrm{old}}$ denotes the behavior policy used to collect the rollout and $\omega_t^{(g)}$ is the current-to-behavior importance ratio for the joint structural action at policy step $t$. TTR uses this ratio to propagate the terminal advantage to the transitions that generated the final sequence. Then, the clipped surrogate is summed over the $H$ policy steps and averaged over the $G$ trajectories:
\begin{equation}
\begin{aligned}
\widehat{\mathcal L}_{\mathrm{TTR}}(\phi;y)
=
-\frac{1}{G}
\sum_{g=1}^{G}\sum_{t=0}^{H-1}
\min\Bigl\{
&\omega_t^{(g)}(\phi)A^{(g)},
\operatorname{clip}
\bigl(
\omega_t^{(g)}(\phi),
1-\epsilon_{\mathrm{clip}},
1+\epsilon_{\mathrm{clip}}
\bigr)A^{(g)}
\Bigr\}.
\end{aligned}
\label{eq:ttr_objective}
\end{equation}
The threshold $\epsilon_{\mathrm{clip}}>0$ defines the clipping interval in the proximal policy optimization (PPO) surrogate~\citep{schulman2017ppo}. To preserve the supervised sequence prior, we add Kullback--Leibler (KL) regularization toward a frozen reference and clean-nucleotide cross-entropy (CE):
\begin{equation}
\mathcal L_{\mathrm{RL}}(\phi)
=
\mathbb E_{y\sim\mathcal D_{\mathrm{RL}}}
\left[
\widehat{\mathcal L}_{\mathrm{TTR}}(\phi;y)
\right]
+
\lambda_{\mathrm{KL}}
\mathcal L_{\mathrm{KL}}(\phi,\phi_{\mathrm{ref}})
+
\lambda_{\mathrm{CE}}
\mathcal L_{\mathrm{CE}}(\phi).
\label{eq:rl_total_objective}
\end{equation}
where $\lambda_{\mathrm{KL}},\lambda_{\mathrm{CE}}\geq0$ weight the KL and CE regularization terms, respectively.
Here $\mathcal D_{\mathrm{RL}}$ denotes the post-training target distribution,
$\phi_{\mathrm{ref}}$ is a frozen reference policy, and
$\phi_{\mathrm{old}}$ remains the behavior policy associated with the current
rollout. $\mathcal L_{\mathrm{KL}}$ regularizes the structured resampling
distributions toward the reference policy, whereas
$\mathcal L_{\mathrm{CE}}$ retains the clean-nucleotide supervision learned by
RNA-IFlow. 

\paragraph{Inference.}

At inference, RNA-IFlow and RNA-IFlow-RL follow different generation dynamics. RNA-IFlow evolves an initial simplex state through global transport via Equation~\ref{eq:iflow_transport} and then applies structure-aware terminal decoding. In contrast, RNA-IFlow-RL uniformly initializes each structural unit over its legal state space and directly generates a discrete sequence through an $H$-step RL policy. For each target, an inference budget of $K$ candidates is realized by $K$ independent trajectories. Thermodynamic feedback is used during post-training, not to guide inference-time sampling or reranking. Generated sequences may subsequently be scored for evaluation.

\section{Experiments}
\label{sec:experiments}

We first evaluate both stages of our framework on standard RNA design benchmarks, then analyze post-training dynamics and the flow-to-policy transition, and finally present a case study.

\subsection{Experimental Setup}
\label{sec:experimental_setup}

\paragraph{Dataset and benchmarks.}
RNA-IFlow is trained on 10 million sequence--structure pairs released by RNA-Design-LM~\citep{gautam2026rnadesignlm}.
These pairs are constructed from one million random RNAs of 6--500 nucleotides by folding each sequence with ViennaRNA\footnote{Official site: \url{https://www.tbi.univie.ac.at/RNA/}}~\citep{lorenz2011vienna} and generating ten sequence designs per target with SAMFEO~\citep{zhou2023samfeo}.
RNA-IFlow-RL uses the released 2,790-target EternaWeb post-training set~\citep{koodli2019eternabrain,gautam2026rnadesignlm}. It excludes structures that exceed 500 nucleotides or cannot be designed by MFE, and further selects targets with moderate design difficulty using the supervised model.
We evaluate on Eterna100-v2, Eterna100, and Rfam-27~\citep{taneda2011modena,andersonlee2016eterna,koodli2021eterna2}, and report additional results on RNAsolo~\citep{adamczyk2022rnasolo} in Appendix~\ref{app:benchmark_scope}.

\paragraph{Baselines.}
\textbf{(1) Conditional language models:}
RNA-Design-LM SL is an autoregressive model trained with supervised learning, while RNA-Design-LM SL+RL further applies RL training to it~\citep{gautam2026rnadesignlm}.
GoForth uses an encoder--decoder architecture for RNA design~\citep{lindsey2026goforth}.
\textbf{(2) Learning-based optimization methods:}
DRAG performs RNA design through hierarchical graph-based nucleotide mutation~\citep{li2025drag}.
\textbf{(3) Search-based methods:}
RNAinverse-pf optimizes sequences toward a target structure using a partition-function objective~\citep{lorenz2011vienna}.
Additional comparisons with NEMO, SAMFEO and FastDesign are reported in Appendix~\ref{app:native_search_60s}.

\paragraph{Evaluation settings.}
The inference budget is $K=8$ ordered candidates per target.
Pass@1 (P@1) and P@8 denote uMFE success for the first candidate and for at least one of the first eight candidates, respectively.
MFE@8 success allows the target structure to be one of multiple MFE folds.
We also report target probability and normalized ensemble defect (NED) for thermodynamic quality, Pair-F1 for structural agreement, and sequence diversity.
All candidates are evaluated with ViennaRNA.

\paragraph{Implementation details.}
RNA-IFlow uses RNAErnie~\citep{wang2024rnaernie} as the bidirectional LM backbone and contains 87.55M parameters, with 86.96M updated during supervised training and 14.49M during RL post-training.
Supervised Flow Matching samples $\alpha\sim\mathcal{U}(1,8)$ and uses 50 integration steps at inference.
RNA-IFlow-RL uses $G=8$ trajectories per target and $H=8$ policy transitions; its terminal reward combines target probability, MFE success, and uMFE success with weights 0.5, 0.25, and 0.25.
Training RNA-IFlow and RNA-IFlow-RL spent 84 and 48 GPU-hours, respectively.
Full settings are provided in Appendix~\ref{app:evaluation_contracts}.

\begin{table}[t]
\centering\small
\setlength{\tabcolsep}{2.8pt}
\caption{
Overall performance. Bold and underlined values denote the best and second-best results, respectively. The standard deviations are reported in Appendix~\ref{tab:main_results_sd}.
}
\label{tab:main_results}
\begin{adjustbox}{width=1\linewidth,center}
\begin{tabular}{llr*{6}{S[table-format=1.4]}}
\toprule
\multicolumn{1}{c}{\multirow[c]{2}{*}{Method}} & \multicolumn{1}{c}{\multirow[c]{2}{*}{Type}} & \multicolumn{1}{c}{\multirow[c]{2}{*}{Params (M)}} & \multicolumn{2}{c}{Eterna100-v2} & \multicolumn{2}{c}{Eterna100} & \multicolumn{2}{c}{Rfam-27}\\
\cmidrule(lr){4-5}\cmidrule(lr){6-7}\cmidrule(lr){8-9}
&&&{P@1 $\uparrow$}&{P@8 $\uparrow$}&{P@1 $\uparrow$}&{P@8 $\uparrow$}&{P@1 $\uparrow$}&{P@8 $\uparrow$}\\
\midrule
RNA-Design-LM SL & Conditional gen. & 357.92 & 0.2633 & 0.4333 & 0.2633 & 0.4267 & 0.5185 & 0.7531\\
RNA-Design-LM SL+RL & Conditional gen. & 357.92 & \second{0.5133} & \second{0.6367} & \second{0.5033} & \second{0.6133} & \second{0.8148} & \second{0.8642}\\
GoForth & Conditional gen. & {59.39} & 0.3933 & 0.5700 & 0.3933 & 0.5667 & 0.6296 & 0.7284\\
\midrule
DRAG & Learned opt. & {0.04} & 0.3633 & 0.5133 & 0.3633 & 0.5133 & 0.7160 & 0.8272\\
RNAinverse-pf & Search & {-} & 0.4367 & 0.5433 & 0.4433 & 0.5100 & 0.3827 & 0.4568\\
\midrule
\textbf{RNA-IFlow (Ours)} & Flow gen. & 87.55 & 0.3033 & 0.5200 & 0.3000 & 0.4867 & 0.6543 & 0.8272\\
\textbf{RNA-IFlow-RL (Ours)} & Flow + RL & 87.55 & \best{0.5400} & \best{0.6500} & \best{0.5067} & \best{0.6167} & \best{0.8519} & \best{0.8765}\\
\bottomrule
\end{tabular}
\end{adjustbox}
\end{table}
\begin{table}[t]
\centering\small
\setlength{\tabcolsep}{2.1pt}
\caption{
Thermodynamic quality and runtime for eight candidates. Lower NED is better. Runtimes are measured on the same machines, while starred rows include internal search and folding.
}
\label{tab:thermodynamic_efficiency}
\begin{adjustbox}{width=\linewidth,center}
\begin{tabular}{lS[table-format=1.4]S[table-format=1.4]S[table-format=1.4]S[table-format=1.4]S[table-format=1.4]S[table-format=1.4]S[table-format=1.4]r}
\toprule
\multicolumn{1}{c}{\multirow[c]{2}{*}{Method}} & \multicolumn{3}{c}{Eterna100-v2} & \multicolumn{2}{c}{Eterna100} & \multicolumn{2}{c}{Rfam-27} & \multicolumn{1}{c}{\multirow[c]{2}{*}{Run time (s)}}\\
\cmidrule(lr){2-4}\cmidrule(lr){5-6}\cmidrule(lr){7-8}
 & {MFE@8 $\uparrow$} & {Prob. $\uparrow$} & {NED $\downarrow$} & {Prob. $\uparrow$} & {NED $\downarrow$} & {Prob. $\uparrow$} & {NED $\downarrow$} & {}\\
\midrule
RNA-Design-LM SL & 0.4433 & 0.3364 & 0.1694 & 0.3326 & 0.1702 & 0.6200 & 0.0267 & 4.41\\
RNA-Design-LM SL+RL & \second{0.6567} & \second{0.5256} & \second{0.0652} & \second{0.5165} & \second{0.0691} & \second{0.7765} & \second{0.0058} & 4.41\\
GoForth & 0.5833 & 0.3499 & 0.0804 & 0.3435 & 0.0848 & 0.4878 & 0.0162 & 1.35\\
DRAG$^{*}$ & 0.5200 & 0.2391 & 0.1345 & 0.2376 & 0.1373 & 0.4218 & 0.0468 & 14.86\\
RNAinverse-pf$^{*}$ & 0.5700 & 0.4344 & 0.3625 & 0.4292 & 0.3594 & 0.4375 & 0.5076 & 133.56\\
\midrule
\textbf{RNA-IFlow (Ours)} & 0.5367 & 0.4089 & 0.1017 & 0.3988 & 0.1099 & 0.7158 & 0.0104 & \underline{0.88}\\
\textbf{RNA-IFlow-RL (Ours)} & \best{0.6733} & \best{0.5398} & \best{0.0534} & \best{0.5283} & \best{0.0593} & \best{0.7785} & \best{0.0057} & \textbf{0.73}\\
\bottomrule
\end{tabular}
\end{adjustbox}
\end{table}

\subsection{Main Results}
\label{sec:main_results}

\paragraph{RNA-IFlow-RL improves RNA design performance.}
As shown in Table~\ref{tab:main_results}, RNA-IFlow already improves P@8 over RNA-Design-LM SL from 0.4333 to 0.5200 on Eterna100-v2, from 0.4267 to 0.4867 on Eterna100, and from 0.7531 to 0.8272 on Rfam-27.
RL post-training further raises P@8 to 0.6500, 0.6167, and 0.8765, respectively.
On Eterna100-v2, RNA-IFlow-RL exceeds RNA-Design-LM SL+RL, DRAG, and RNAinverse-pf by 1.33, 13.67, and 10.67 percentage points in P@8, while also achieving the highest target probability and lowest NED across the three benchmarks (Table~\ref{tab:thermodynamic_efficiency}).
These results show that RNA-IFlow-RL consistently improves both design success and thermodynamic quality across benchmarks.

\paragraph{RNA-IFlow-RL is parameter-efficient and fast at inference.}
RNA-IFlow-RL uses 87.55M parameters, compared with 357.92M for RNA-Design-LM, while achieving stronger design performance.
Moreover, Table~\ref{tab:thermodynamic_efficiency} shows that RNA-IFlow-RL generates eight candidates in 0.73\,s, compared with 1.35\,s for GoForth and 4.41\,s for RNA-Design-LM SL+RL. 
These results demonstrate that RNA-IFlow-RL is efficient in both model size and inference speed.

\begin{figure}[!t]
\centering
\includegraphics[width=0.9\textwidth]{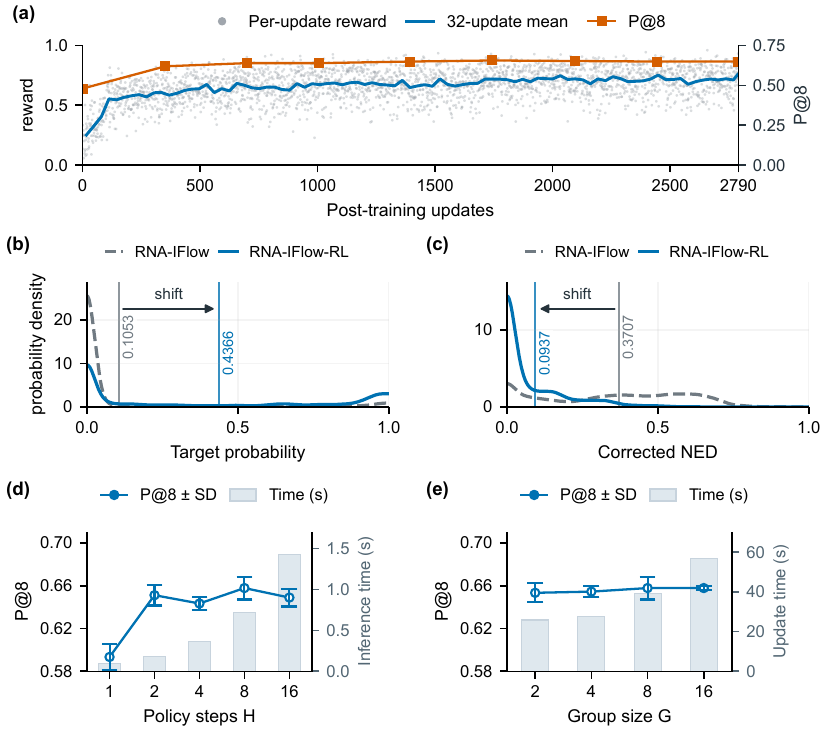}
\caption{
Post-training dynamics and sensitivity analysis. (a) shows reward and P@8 changes during post-training. (b,c) show shifts in target-probability and NED distributions. (d,e) show mean P@8 $\pm$ SD across three seeds for $H$ and $G$, together with inference and update times, respectively.
}
\label{fig:posttraining_hg}
\end{figure}

\subsection{Post-Training Dynamics and Robustness}
\label{sec:refinement_scaling}

\paragraph{Post-training converges and shifts the thermodynamic quality distribution.}
Figure~\ref{fig:posttraining_hg} (a) shows the stabilization of RL post-training: after 1500 updates, the terminal reward reaches a plateau and P@8 follows the same overall trend. Figure~\ref{fig:posttraining_hg}(b,c) further shows the thermodynamic distribution shift induced by RL: RNA-IFlow-RL shifts target probability toward higher values and NED toward lower values compared with RNA-IFlow.

\paragraph{Performance is stable across policy steps and group sizes.}
As shown in Figure~\ref{fig:posttraining_hg}(d,e), P@8 remains stable across the tested $H$ and $G$ values, while inference and update time increase with the policy horizon $H$ and the group size $G$, respectively.
Therefore, the setting $H=8$ and $G=8$ provides a trade-off between design quality and time cost. 
More analyses are shown in Appendix~\ref{app:current_sensitivity}.

\begin{figure}[!t]
\centering
\includegraphics[width=0.9\textwidth]{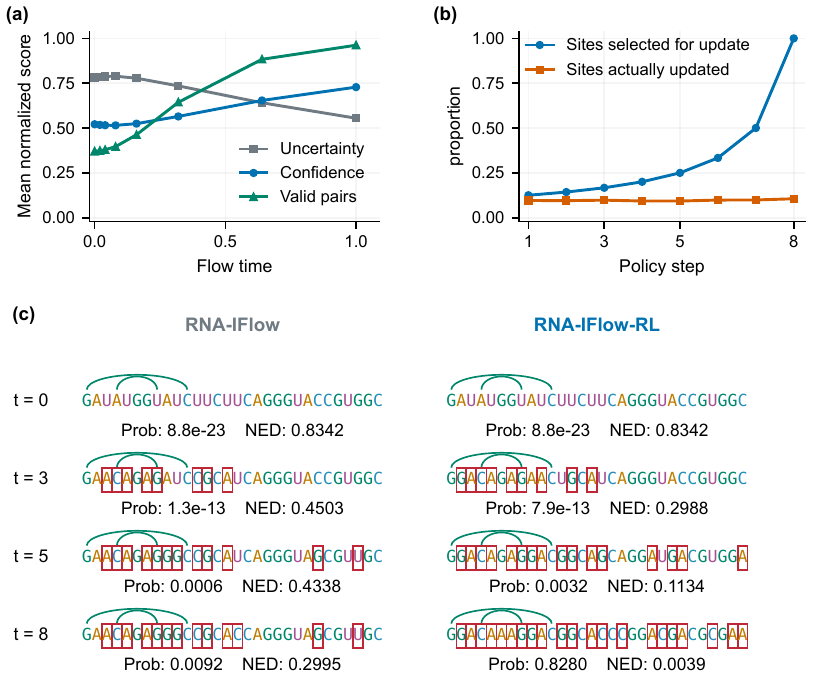}
\caption{
Continuous-flow and finite-policy sequence evolution. 
(a) shows sequence uncertainty, prediction confidence, and the proportion of legal target pairs over normalized flow time. 
(b) compares scheduled (blue) and actual (orange) nucleotide updates across policy steps. (c) compares matched RNA-IFlow and RNA-IFlow-RL trajectories; red boxes mark differences from the initial sequence.
}
\label{fig:flow_to_policy}
\end{figure}
\subsection{From Continuous Flow to a Finite RL Policy}
\label{sec:flow_policy_analysis}

Figure~\ref{fig:flow_to_policy} (a) shows how RNA-IFlow gradually transforms an uncertain sequence state into a more confident and pairing-compatible sequence along the continuous flow.
After converting the flow predictor into a finite policy, more sites are considered for variation at later steps, while only a small fraction of nucleotides are actually changed (Figure~\ref{fig:flow_to_policy} (b)).
This indicates that the policy performs selective sequence refinement rather than simply rewriting the whole sequence.
In Figure~\ref{fig:flow_to_policy} (c), the flow model and RL policy start from the same sequence, but RNA-IFlow-RL consistently makes beneficial variations with substantially higher target probability and lower NED.
This demonstrates that RL training guides the policy to learn which sequence variations are thermodynamically useful.

\subsection{Case Study}
\label{sec:case_study}

\begin{figure}[htbp]
\centering
\includegraphics[width=\textwidth]{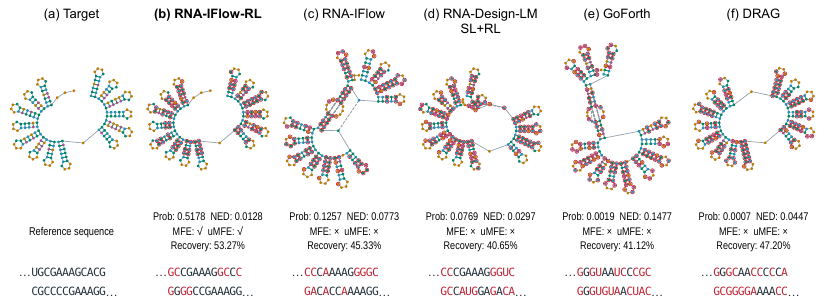}
\caption{This figure compares Anemone designs (214 nt). Scores and recovery use full sequences to evaluate. In the visualization of secondary structure, red bases mark differences from the reference, while green, orange, and dashed gray bonds denote recovered, alternative, and missing pairs.}
\label{fig:structure_refinement}
\end{figure}

On the 214-nt Anemone target, the displayed RNA-IFlow-RL candidate reaches target probability 0.5178, NED 0.0128, and uMFE success, whereas the displayed candidates from the comparison methods do not achieve target MFE success (Figure~\ref{fig:structure_refinement}).
RNA-IFlow-RL also reaches 53.27\% sequence recovery to the player-designed reference, compared with 40.65--47.20\% for the other methods.
The case study visually demonstrates the superiority of RNA-IFlow-RL in RNA design.

\section{Conclusion}
\label{sec:conclusion}

In this work, we present a two-stage framework comprising RNA-IFlow and RNA-IFlow-RL for RNA design, combining structure-conditioned Dirichlet FM with thermodynamic policy refinement.
The flow model captures coordinated sequence-wide variation, while the finite policy uses terminal folding feedback to shift generation toward thermodynamically favorable sequences. 
Extensive experiments show that RNA-IFlow-RL outperforms comparative methods across multiple benchmarks. 
Further analyses reveal thermodynamic gains, policy dynamics, and robustness in our framework.
These findings demonstrate that our framework advances RNA design and opens a promising direction for broader RNA design modeling.
\label{sec:main_text_end}
\clearpage
\section*{Ethics statement}
This study concerns computational RNA design. Biological use of the designed sequences requires experimental validation and appropriate biosafety review.
\section*{Reproducibility statement}
The main text defines the model, objectives, and evaluation protocols. The appendix records the training data, candidate budgets, decoder settings, sensitivity studies, search configurations, timing scopes, and case-selection rules. Tables and figures are generated from versioned full-precision results.
The structural case uses unchanged full-length sequences and a common cropped display window. 
Code and reproduction scripts are available at \url{https://github.com/John-Lin98/RNA-IFlow}.
\section*{AI use statement}
Generative AI tools were used to assist literature retrieval, research ideation and experimental workflows, manuscript drafting and revision, software review, and figure preparation. All AI-assisted research outputs, code, experimental results, citations, and interpretations were reviewed and verified by the authors. All reported numerical results originate from recorded model evaluations and RNA folding calculations. The authors take full responsibility for the final content.
\bibliography{references}

@article{lorenz2011vienna,
  title={{ViennaRNA} Package 2.0},
  author={Lorenz, Ronny and Bernhart, Stephan H. and H{\"o}ner zu Siederdissen, Christian and Tafer, Hakim and Flamm, Christoph and Stadler, Peter F. and Hofacker, Ivo L.},
  journal={Algorithms for Molecular Biology}, volume={6}, number={1}, pages={26}, year={2011},
  doi={10.1186/1748-7188-6-26}
}

@article{andersonlee2016eterna,
  title={Principles for Predicting {RNA} Secondary Structure Design Difficulty},
  author={Anderson-Lee, Jeff and Fisker, Eli and Kosaraju, Vineet and Wu, Michelle and Kong, Justin and Lee, Jeehyung and Lee, Minjae and Zada, Mathew and Treuille, Adrien and Das, Rhiju and {Eterna Players}},
  journal={Journal of Molecular Biology}, volume={428}, number={5, Part A}, pages={748--757}, year={2016},
  doi={10.1016/j.jmb.2015.11.013}
}

@article{fieldhouse2026moderna,
  title   = {Moderna cancer vaccine stops melanoma returning: what’s next for personalized treatments?},
  author  = {Fieldhouse, Rachel and Basu, Mohana},
  journal = {Nature},
  volume  = {657},
  pages   = {16--17},
  year    = {2026},
  doi     = {10.1038/d41586-026-02612-3},
  url     = {https://doi.org/10.1038/d41586-026-02612-3}
}

@article{taneda2011modena,
  title={{MODENA}: A Multi-objective {RNA} Inverse Folding}, author={Taneda, Akito},
  journal={Advances and Applications in Bioinformatics and Chemistry}, volume={4}, pages={1--12}, year={2011},
  doi={10.2147/AABC.S14335}
}

@article{andronescu2004rnassd,
  title={A New Algorithm for {RNA} Secondary Structure Design},
  author={Andronescu, Mirela and Fejes, Anthony P. and Hutter, Frank and Hoos, Holger H. and Condon, Anne},
  journal={Journal of Molecular Biology}, volume={336}, number={3}, pages={607--624}, year={2004},
  doi={10.1016/j.jmb.2003.12.041}
}

@article{zadeh2011nupack,
  title={{NUPACK}: Analysis and Design of Nucleic Acid Systems},
  author={Zadeh, Joseph N. and Steenberg, Conrad D. and Bois, Justin S. and Wolfe, Brian R. and Pierce, Marshall B. and Khan, Asif R. and Dirks, Robert M. and Pierce, Niles A.},
  journal={Journal of Computational Chemistry}, volume={32}, number={1}, pages={170--173}, year={2011},
  doi={10.1002/jcc.21596}
}

@article{koodli2019eternabrain,
  title={{EternaBrain}: Automated {RNA} Design through Move Sets and Strategies from an Internet-scale {RNA} Videogame},
  author={Koodli, Rohan V. and Keep, Benjamin and Coppess, Katherine R. and Portela, Fernando and {Eterna participants} and Das, Rhiju},
  journal={PLoS Computational Biology}, volume={15}, number={6}, pages={e1007059}, year={2019},
  doi={10.1371/journal.pcbi.1007059}
}

@article{zhou2023samfeo,
  title={{RNA} Design via Structure-Aware Multifrontier Ensemble Optimization},
  author={Zhou, Tianshuo and Dai, Ning and Li, Sizhen and Ward, Max and Mathews, David H. and Huang, Liang},
  journal={Bioinformatics}, volume={39}, number={Supplement 1}, pages={i563--i571}, year={2023},
  doi={10.1093/bioinformatics/btad252}
}

@article{ward2023fitness,
  title={Fitness Functions for {RNA} Structure Design},
  author={Ward, Max and Courtney, Eliot and Rivas, Elena},
  journal={Nucleic Acids Research}, volume={51}, number={7}, pages={e40}, year={2023},
  doi={10.1093/nar/gkad097}
}

@inproceedings{runge2019learna,
  title={Learning to Design {RNA}}, author={Runge, Frederic and Stoll, Danny and Falkner, Stefan and Hutter, Frank},
  booktitle={International Conference on Learning Representations}, year={2019},
  url={https://openreview.net/forum?id=ByfyHh05tQ}
}

@article{li2025drag,
  title={{DRAG}: Design {RNAs} as Hierarchical Graphs with Reinforcement Learning},
  author={Li, Yichong and Pan, Xiaoyong and Shen, Hongbin and Yang, Yang},
  journal={Briefings in Bioinformatics}, volume={26}, number={2}, pages={bbaf106}, year={2025},
  doi={10.1093/bib/bbaf106},
  url={https://doi.org/10.1093/bib/bbaf106}
}

@article{tang2026samplingdesign,
  title={{SamplingDesign}: {RNA} Design via Continuous Optimization with Coupled Variables and {Monte-Carlo} Sampling},
  author={Tang, Wei Yu and Dai, Ning and Zhou, Tianshuo and Mathews, David H. and Huang, Liang},
  journal={Nature Communications}, volume={17}, pages={2950}, year={2026},
  doi={10.1038/s41467-025-67901-3},
  url={https://doi.org/10.1038/s41467-025-67901-3}
}

@misc{gautam2026rnadesignlm,
  title={Designing {RNAs} with Language Models},
  author={Gautam, Milan and Dai, Ning and Zhou, Tianshuo and Xie, Bowen and Mathews, David and Huang, Liang},
  year={2026}, eprint={2602.12470}, archivePrefix={arXiv}, primaryClass={cs.LG},
  doi={10.48550/arXiv.2602.12470},
  url={https://arxiv.org/abs/2602.12470}
}

@misc{lindsey2026goforth,
  title={{GoForth}: Language Models for {RNA} Design under Structure, Sequence, and Coding Constraints},
  author={Lindsey, Michael}, year={2026}, eprint={2605.07608}, archivePrefix={arXiv}, primaryClass={q-bio.QM},
  doi={10.48550/arXiv.2605.07608}, url={https://arxiv.org/abs/2605.07608}
}

@inproceedings{gat2024dfm,
  title={Discrete Flow Matching},
  author={Gat, Itai and Remez, Tal and Shaul, Neta and Kreuk, Felix and Chen, Ricky T. Q. and Synnaeve, Gabriel and Adi, Yossi and Lipman, Yaron},
  booktitle={Advances in Neural Information Processing Systems}, volume={37}, pages={133345--133385}, year={2024},
  url={https://proceedings.neurips.cc/paper_files/paper/2024/hash/f0d629a734b56a642701bba7bc8bb3ed-Abstract-Conference.html}
}

@inproceedings{lipman2023flow,
  title={Flow Matching for Generative Modeling},
  author={Lipman, Yaron and Chen, Ricky T. Q. and Ben-Hamu, Heli and Nickel, Maximilian and Le, Matt},
  booktitle={International Conference on Learning Representations}, year={2023},
  url={https://openreview.net/forum?id=PqvMRDCJT9t}
}

@misc{schulman2017ppo,
  title={Proximal Policy Optimization Algorithms},
  author={Schulman, John and Wolski, Filip and Dhariwal, Prafulla and Radford, Alec and Klimov, Oleg},
  year={2017}, eprint={1707.06347}, archivePrefix={arXiv}, primaryClass={cs.LG},
  doi={10.48550/arXiv.1707.06347},
  url={https://arxiv.org/abs/1707.06347}
}

@inproceedings{nori2024rnaflow,
  title={{RNAFlow}: {RNA} Structure \& Sequence Design via Inverse Folding-Based Flow Matching},
  author={Nori, Divya and Jin, Wengong},
  booktitle={Proceedings of the 41st International Conference on Machine Learning},
  series={Proceedings of Machine Learning Research}, volume={235},
  pages={38395--38408}, year={2024}, publisher={PMLR},
  url={https://proceedings.mlr.press/v235/nori24a.html}
}

@article{adamczyk2022rnasolo,
  title={{RNAsolo}: A Repository of Cleaned {PDB}-Derived {RNA} 3{D} Structures},
  author={Adamczyk, Bartosz and Antczak, Maciej and Szachniuk, Marta},
  journal={Bioinformatics}, volume={38}, number={14}, pages={3668--3670}, year={2022},
  doi={10.1093/bioinformatics/btac386},
  url={https://doi.org/10.1093/bioinformatics/btac386}
}

@misc{zhou2026fastdesign,
  title={Fast and Versatile {RNA} Design via Motif-Level Divide-and-Conquer and Structure-Level Rival Search},
  author={Zhou, Tianshuo and Mathews, David H. and Huang, Liang},
  year={2026}, eprint={2603.02283}, archivePrefix={arXiv}, primaryClass={q-bio.BM},
  doi={10.48550/arXiv.2603.02283},
  url={https://arxiv.org/abs/2603.02283}
}

@inproceedings{mcallister2026fpo,
  title={Flow Matching Policy Gradients},
  author={McAllister, David and Ge, Songwei and Yi, Brent and Kim, Chung Min and Weber, Ethan and Choi, Hongsuk and Feng, Haiwen and Kanazawa, Angjoo},
  booktitle={International Conference on Learning Representations},
  year={2026},
  url={https://proceedings.iclr.cc/paper_files/paper/2026/hash/3d43cc5692bf68944ee7cd31b97d0c11-Abstract-Conference.html}
}

@misc{su2026domino,
  title={Discrete Flow Matching Policy Optimization},
  author={Su, Maojiang and Hsieh, Po-Chung and Wu, Weimin and Lu, Mingcheng and Chen, Jiunhau and Hu, Jerry Yao-Chieh and Liu, Han},
  year={2026}, eprint={2604.06491}, archivePrefix={arXiv}, primaryClass={cs.LG},
  doi={10.48550/arXiv.2604.06491},
  url={https://arxiv.org/abs/2604.06491}
}

@inproceedings{wang2025drakes,
  title={Fine-Tuning Discrete Diffusion Models via Reward Optimization with Applications to {DNA} and Protein Design},
  author={Wang, Chenyu and Uehara, Masatoshi and He, Yichun and Wang, Amy and Lal, Avantika and Jaakkola, Tommi and Levine, Sergey and Regev, Aviv and Wang, Hanchen and Biancalani, Tommaso},
  booktitle={International Conference on Learning Representations},
  year={2025}, eprint={2410.13643}, archivePrefix={arXiv},
  url={https://proceedings.iclr.cc/paper_files/paper/2025/hash/771e09dd204ea339da0d8114c48afd21-Abstract-Conference.html}
}

@inproceedings{ma2025riboflow,
  title={{RiboFlow}: Conditional De Novo {RNA} Co-Design via Synergistic Flow Matching},
  author={Ma, Runze and Zhang, Zhongyue and Wang, Zichen and Hua, Chenqing and Rao, Jiahua and Zhou, Zhuomin and Zheng, Shuangjia},
  booktitle={Advances in Neural Information Processing Systems},
  volume={38}, pages={77810--77842}, year={2025},
  doi={10.52202/085713-2348},
  url={https://proceedings.neurips.cc/paper_files/paper/2025/hash/653353d903d87720d781a9554b0018db-Abstract-Conference.html}
}

@inproceedings{stark2024dirichlet,
  title={Dirichlet Flow Matching with Applications to {DNA} Sequence Design},
  author={Stark, Hannes and Jing, Bowen and Wang, Chenyu and Corso, Gabriele and Berger, Bonnie and Barzilay, Regina and Jaakkola, Tommi},
  booktitle={Proceedings of the 41st International Conference on Machine Learning},
  series={Proceedings of Machine Learning Research}, volume={235},
  pages={46495--46513}, year={2024}, publisher={PMLR},
  url={https://proceedings.mlr.press/v235/stark24b.html}
}

@article{dutheil2010epistasis,
  title={Base Pairing Constraints Drive Structural Epistasis in Ribosomal {RNA} Sequences},
  author={Dutheil, Julien Y. and Jossinet, Fabrice and Westhof, Eric},
  journal={Molecular Biology and Evolution}, volume={27}, number={8},
  pages={1868--1876}, year={2010}, doi={10.1093/molbev/msq069}
}

@article{chen1999compensatory,
  title={{RNA} Secondary Structure and Compensatory Evolution},
  author={Chen, Ying and Carlini, David B. and Baines, John F. and Parsch, John and Braverman, John M. and Tanda, Soichi and Stephan, Wolfgang},
  journal={Genes \& Genetic Systems}, volume={74}, number={6},
  pages={271--286}, year={1999}, doi={10.1266/ggs.74.271}
}

@article{koodli2021eterna2,
  title={Redesigning the {Eterna100} for the {Vienna 2} Folding Engine},
  author={Koodli, Rohan V. and Rudolfs, Boris and Wayment-Steele, Hannah K. and {Eterna Structure Designers} and Das, Rhiju},
  journal={bioRxiv}, year={2021},
  doi={10.1101/2021.08.26.457839},
  url={https://www.biorxiv.org/content/10.1101/2021.08.26.457839v1}
}

@article{portela2018nemo,
  title={An Unexpectedly Effective {Monte Carlo} Technique for the {RNA} Inverse Folding Problem},
  author={Portela, Fernando}, journal={bioRxiv}, year={2018},
  doi={10.1101/345587},
  url={https://www.biorxiv.org/content/10.1101/345587v1}
}

@article{wang2024rnaernie,
  title={Multi-purpose {RNA} Language Modelling with Motif-aware Pretraining and Type-guided Fine-tuning},
  author={Wang, Ning and Bian, Jiang and Li, Yuchen and Li, Xuhong and Mumtaz, Shahid and Kong, Linghe and Xiong, Haoyi},
  journal={Nature Machine Intelligence}, volume={6}, pages={548--557}, year={2024},
  doi={10.1038/s42256-024-00836-4},
  url={https://www.nature.com/articles/s42256-024-00836-4}
}
\bibliographystyle{iclr2027_conference}

\clearpage
\appendix
\section{Additional Methods and Experiments}
The supplementary material records the evaluation contracts, complete benchmark results, policy and data controls, native search protocols, large-budget references, and structural examples. All model variants retain their recorded checkpoints and generation interfaces.
\subsection{Extended sampling and data analyses}
\label{app:extended_analysis}

\begin{figure}[hb]
\centering
\includegraphics[width=0.9\textwidth]{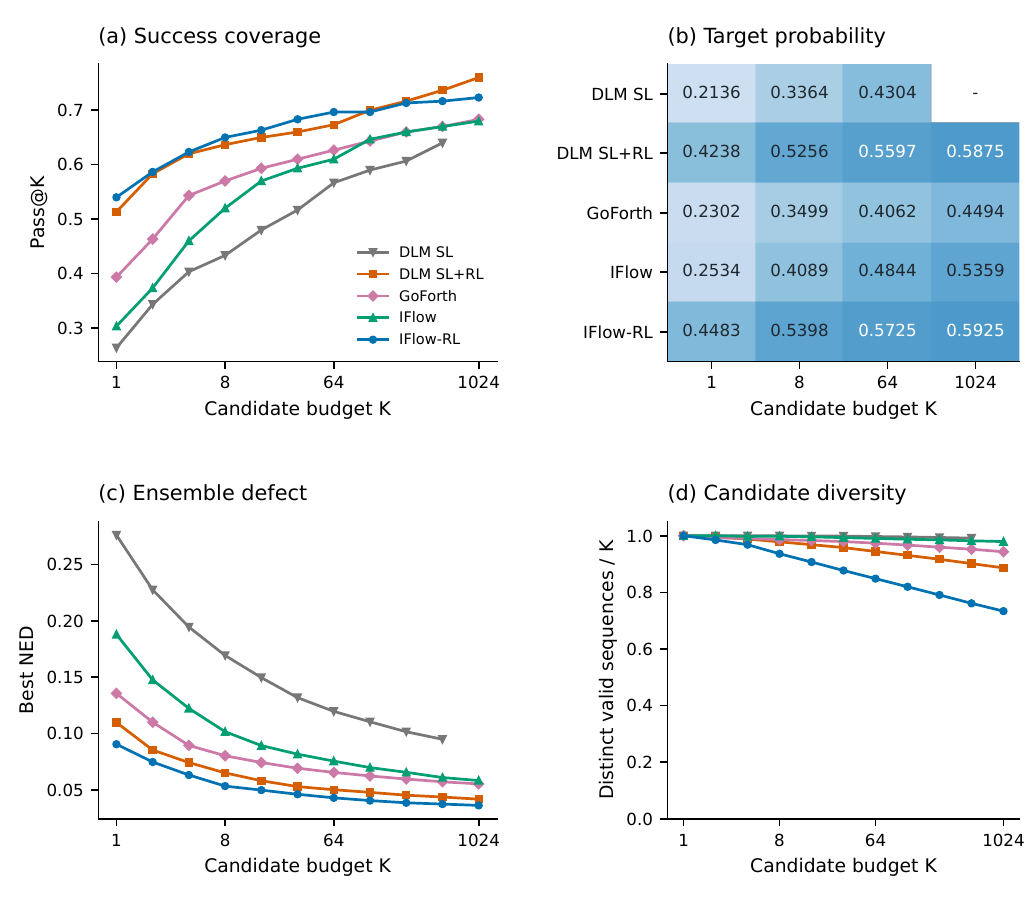}
\caption{This figure compares candidate budgets on Eterna100-v2. Panel (a) shows uMFE coverage, panels (b,c) show the best target probability and NED, and panel (d) shows distinct valid sequences. Curves use nested candidate prefixes.}
\label{fig:budget_difficulty_runtime}
\end{figure}

\paragraph{Candidate budget reveals a quality--diversity trade-off.}
RNA-IFlow-RL has higher Pass@$K$ than RNA-Design-LM SL+RL through $K=64$, whereas RNA-Design-LM SL+RL becomes higher at larger budgets (Figure~\ref{fig:budget_difficulty_runtime}a).
RNA-IFlow-RL nevertheless retains higher target probability and lower NED across the measured prefixes, while its fraction of distinct sequences decreases more rapidly as $K$ grows (Figure~\ref{fig:budget_difficulty_runtime}b--d).
Thus, sampling budget improves coverage and thermodynamic quality at different rates.

\begin{figure}[tbp]
\centering
\includegraphics[width=0.9\textwidth]{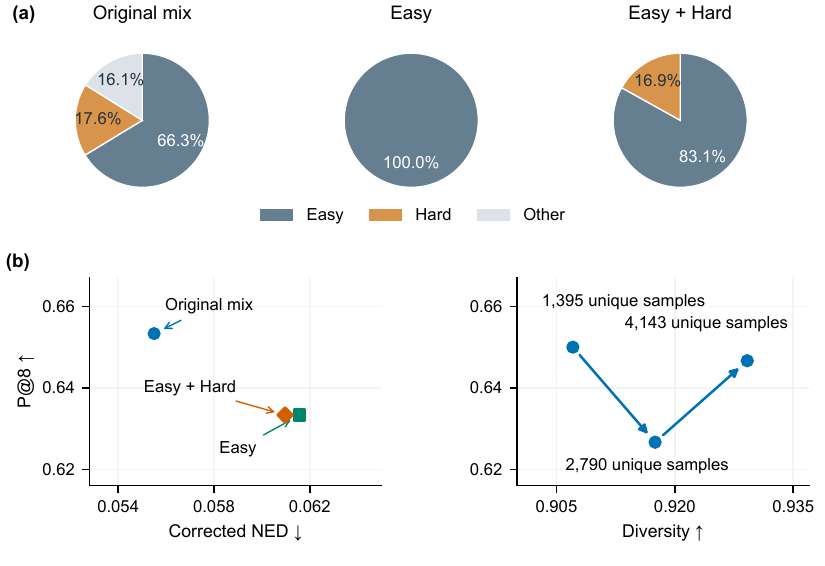}
\caption{This figure analyzes training-data composition and coverage. Panel (a) compares equal-size training mixtures, while panel (b) relates composition to quality (left) and fixed-exposure coverage to diversity (right). Unique samples denote distinct target structures.}
\label{fig:training_composition}
\end{figure}

\paragraph{Training composition and coverage affect quality and diversity differently.}
At matched training-set size, the Original mix achieves higher P@8 and lower NED than the restricted Easy and Easy + Hard sets (Figure~\ref{fig:training_composition}a,b).
Under a fixed number of target visits, increasing the number of unique training samples raises sequence diversity, while P@8 changes non-monotonically (Figure~\ref{fig:training_composition}b).
This exposes a coverage--revisitation trade-off rather than a simple scaling law.

\begin{figure}[tbp]
\centering
\includegraphics[width=0.9\textwidth]{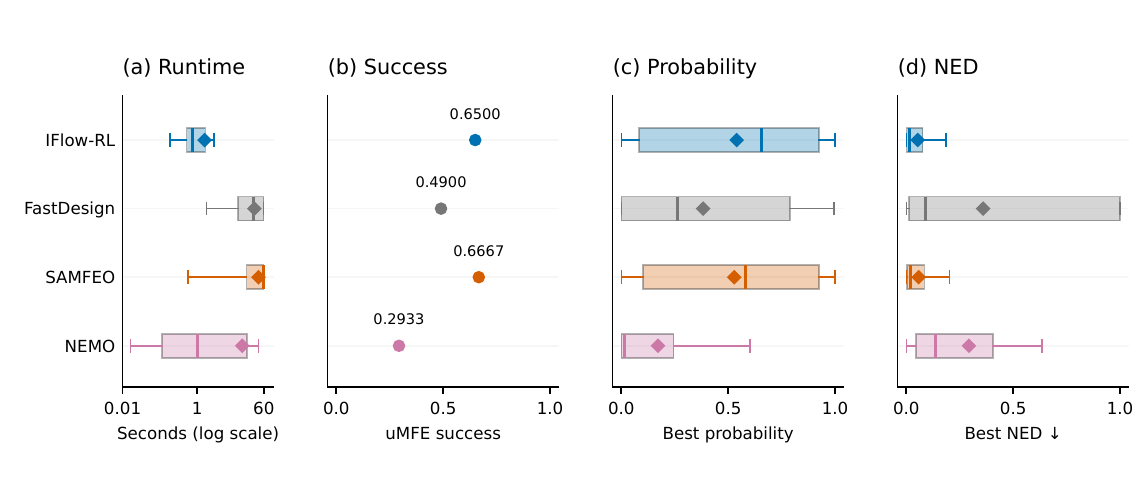}
\caption{This figure compares quality and cost under the recorded generation and native-search protocols. Panel (a) shows runtime, panel (b) shows uMFE success, and panels (c,d) show target probability and NED. Boxes show quartiles and diamonds denote means.}
\label{fig:search_frontier}
\end{figure}

\paragraph{Search and neural generation occupy different quality--cost regimes.}
Under a native 60-second search limit, SAMFEO attains slightly higher uMFE success than RNA-IFlow-RL, whereas RNA-IFlow-RL achieves higher target probability and lower NED (Figure~\ref{fig:search_frontier}).
The comparison therefore reflects different quality--compute trade-offs rather than a single metric advantage.
Full native-search and large-budget results are reported in Appendices~\ref{app:native_search_60s} and~\ref{app:published_provenance}.

\FloatBarrier

\subsection{Figure guide and evaluation scope}
\label{app:figure_guide}
\paragraph{Method overview.}
Figure~\ref{fig:method_overview} illustrates coordinated variation under a fixed pairing topology, supervised RNA-IFlow training, and RNA-IFlow-RL post-training. In Stage 1, the loss is computed from the clean-nucleotide predictor before terminal decoding. The separate generation branch transports the continuous state and decodes it into a sequence. In Stage 2, terminal folding scores produce group-relative advantages for policy updates. Flames denote trainable parameters and snowflakes denote frozen inference parameters. Highlighted bases show changes from the preceding displayed state. Sequences and folding graphics in the overview are schematic.

\paragraph{Post-training distributions and sensitivity.}
Figure~\ref{fig:posttraining_hg}(a) summarizes terminal reward and P@8 during post-training.
Panels (b,c) show the candidate-level distributions of target probability and NED, with solid vertical lines marking the corresponding means.
Panels (d,e) report sensitivity to the policy horizon $H$ and group size $G$.
The $G=1$ setting has zero group-relative advantage and is listed separately in Table~\ref{tab:hg_sensitivity}.

\paragraph{From continuous flow to a finite policy.}
Figure~\ref{fig:flow_to_policy}(a) summarizes native continuous-flow trajectories over 13 targets. Uncertainty, confidence, and the legal-pair fraction use discrete site-wise predictor readouts. Panel (b) compares the finite-policy schedule's expected site-selection rate with the observed nucleotide-change rate in post-trained trajectories. Prob. and NED evaluate the full sequence after sampling; the displayed window is used only for visualization.

\paragraph{Training mixture and coverage.}
Figure~\ref{fig:training_composition}(a) shows three equal-size training sets. The Original mix contains 66.3\% Easy, 17.6\% Hard, and 16.1\% Other samples. Easy is selected using the predefined learnability criterion and contains 100\% Easy samples; Easy + Hard contains 83.1\% Easy and 16.9\% Hard samples. These latter sets correspond to the core-only and hard-enriched names in Table~\ref{tab:composition_corrected}. In panel (b), the left plane compares their NED and P@8, and the right plane compares diversity and P@8 under 4,144 target visits. Unique samples denote distinct target structures. Every configuration uses one training seed, and broader coverage reduces the mean number of repeated visits.

\paragraph{Case interpretation.}
Figure~\ref{fig:structure_refinement} compares the Anemone target and five methods. The displayed sequences show positions 83--132, while structures, scores, and recovery use all 214 nucleotides. Recovery is sequence identity to the Eterna player-designed reference. Red letters and node outlines indicate reference-sequence differences; structural agreement is encoded separately by green recovered pairs, orange alternative pairs, and dashed gray missing pairs. MFE and uMFE marks refer to the displayed candidate. Table~\ref{tab:case_metrics} supplies group-level hit counts and Appendix~\ref{app:five_method_case} records the candidate-selection rule.

\paragraph{Result version and training cost.}
The numerical tables and author-review figures in this manuscript report results across three evaluation seeds. The supervised training cost of 84.1 A100 GPU-hours covers eight epochs, with epoch 6 selected for evaluation. The complete RL training lineage used to obtain the reported model requires approximately 48.0 allocated A100 GPU-hours.

\subsection{Theory and implementation alignment}
\label{app:theory}

\paragraph{Index and distribution conventions.}
$g$ labels an RL trajectory, $t\in\{0,\ldots,H\}$ its state index, $j$ a nucleotide position, and $u\in\mathcal U(y)$ a structural unit. Thus $x_{t,j}^{(g)}$ is one nucleotide, $x_t^{(g)}$ the complete sequence, and $z_t^{(g)}$ its simplex representation. Static $x_j$ denotes the clean nucleotide in a sequence--structure training example. $G$ is the number of training trajectories per target, $H$ the policy horizon, and $K$ the evaluation candidate budget. Free energy is $\mathcal G$, distinct from $G$. The model uses parameters $\theta$ before post-training and $\phi$ during refinement; $\phi_{\mathrm{old}}$ identifies the behavior model and $\phi_{\mathrm{ref}}$ the fixed reference.

\paragraph{Conditional-flow coefficient and numerical transport.}
For $\xi\in(0,1)$, the four-category Dirichlet conditional-flow coefficient is
\begin{equation}
c_\alpha(\xi)=-\frac{\partial I_\xi(\alpha,3)}{\partial\alpha}
\frac{B(\alpha,3)}{(1-\xi)^3\xi^{\alpha-1}},
\label{eq:conditional_flow_coefficient}
\end{equation}
where $B$ is the beta function and $I_\xi$ the regularized incomplete beta function. Equation~\ref{eq:posterior_velocity} averages $c_\alpha(z_{j,a})(e(a)-z_j)$ under the clean-nucleotide predictor. The tangent projection is $\mathsf P_{\mathrm{tan}}(v)=v-\frac14(\sum_a v_a)\mathbf1$ and enforces zero coordinate sum. It is distinct from the Euclidean state projection $\Pi_{\Delta^3}$ used after each numerical Euler update:
\begin{equation}
z_j(\alpha_{k+1})=\Pi_{\Delta^3}\!\left[
z_j(\alpha_k)+(\alpha_{k+1}-\alpha_k)v_{\theta,j}(z(\alpha_k),\alpha_k,y)\right].
\label{eq:numerical_flow_update}
\end{equation}
Here $k$ indexes integration steps. The native-flow evaluation initializes from $\operatorname{Dir}(\mathbf1)$ and uses 50 steps on the numerical grid from $1.001$ to $8.0$. The implementation evaluates the beta derivative through interpolation and guards non-finite values. Simplex projection preserves nonnegativity and unit sum after the tangent-projected update.

\paragraph{Exact conditional-mean bridge.}
The concentration vector of $q_\alpha(\cdot\mid a)$ sums to $\alpha+3$. A Dirichlet vector has mean equal to its concentration vector divided by that sum, proving Equation~\ref{eq:fpm_mean_bridge}. The clean coordinate has mean $\alpha/(\alpha+3)$ and the other coordinates $1/(\alpha+3)$. This identity specifies the bridge, without asserting equality between native continuous transport and the constructed finite-policy dynamics.

\paragraph{Sampling temperature and structural distributions.}
For a fixed trajectory temperature $\vartheta>0$, define
\begin{equation}
\widetilde p_{\phi,j}(a\mid z,\alpha,y)
=\operatorname{Softmax}\!\left(f_\phi(z,\alpha,y)_j/\vartheta\right)_a.
\label{eq:temperature_posterior}
\end{equation}
The untempered predictor is recovered at $\vartheta=1$. Paired-unit probabilities are the normalized products in Equation~\ref{eq:spd_pair_policy}. The same temperature is used for behavior scoring and current-policy rescoring of a trajectory. Policy and unit-distribution notation suppresses this fixed setting.

\paragraph{Normalization and all-step pairing legality.}
Each $h_{\phi,u}$ is normalized on its four- or six-state alphabet. Since $0<\rho_t\leq1$, Equation~\ref{eq:fpm_transition} is a convex mixture of two normalized distributions. Paired units are initialized within $\mathcal A_{\mathrm{pair}}$. If a paired state is legal at step $t$, both keeping it and resampling from $h_{\phi,u}$ retain legality; induction therefore establishes legality through step $H$. The final step has $\rho_{H-1}=1$. This guarantees allowed target-pair states, not that the target is the MFE or uMFE fold.

\paragraph{Finite trajectory probability.}
Let $U_{\mathrm{s}}$ and $U_{\mathrm{p}}$ be the counts of single-position and paired units. The initialization distribution assigns $\nu_y(x_0)=4^{-U_{\mathrm{s}}}6^{-U_{\mathrm{p}}}$ to every legal initial state and zero otherwise. Conditional on the complete preceding state, structural units are sampled independently. Consequently, for a legal trajectory,
\begin{equation}
P_\phi(\tau^{(g)}\mid y)
=\nu_y(x_0^{(g)})\prod_{t=0}^{H-1}
\prod_{u\in\mathcal U(y)}
\pi_{\phi,u}(x_{t+1,u}^{(g)}\mid x_t^{(g)},\alpha_t,y).
\label{eq:trajectory_likelihood}
\end{equation}
Here $P_\phi$ denotes a distribution over complete finite-policy trajectories. The common initialization is independent of $\phi$. Training clips the step ratios defined in Equation~\ref{eq:ttr_ratio}.

\paragraph{Loss reductions and anchors.}
For one target group, $\bar r_y=G^{-1}\sum_g r^{(g)}$ and $\sigma_y^2=G^{-1}\sum_g(r^{(g)}-\bar r_y)^2$. The implementation uses $\epsilon_{\mathrm{adv}}=10^{-4}$ and assigns zero advantages when the reward range is below $10^{-6}$. Advantages and behavior probabilities are held fixed during the corresponding optimizer update. Equation~\ref{eq:ttr_objective} averages over the $G$ trajectories and sums over $H$ transitions; replacing the sum by a time mean would change its scale relative to the anchors.

For $c_t^{(g)}=(z_t^{(g)},\alpha_t,y)$, the KL anchor for one target group is
\begin{equation}
\widehat{\mathcal L}_{\mathrm{KL}}(\phi;y)
=\frac{1}{GH|\mathcal U(y)|}\sum_{g=1}^G\sum_{t=0}^{H-1}\sum_{u\in\mathcal U(y)}
D_{\mathrm{KL}}\!\left(h_{\phi,u}(\cdot\mid c_t^{(g)})\,\Vert\,
 h_{\phi_{\mathrm{ref}},u}(\cdot\mid c_t^{(g)})\right).
\label{eq:kl_anchor}
\end{equation}
The notation $h(\cdot\mid c)$ abbreviates the explicitly conditioned distribution in the main text. Averaging this group loss over training targets gives $\mathcal L_{\mathrm{KL}}$. It compares the current and reference \emph{resampling distributions}. For a supervised batch $\mathcal B$ of clean sequence--structure pairs with sampled Dirichlet states, the CE anchor is the valid-token mean
\begin{equation}
\widehat{\mathcal L}_{\mathrm{CE}}(\phi;\mathcal B)
=-\frac{\sum_{(x,y)\in\mathcal B}\sum_{j=1}^{|x|}\log p_{\phi,j}(x_j\mid z,\alpha,y)}
{\sum_{(x,y)\in\mathcal B}|x|}.
\label{eq:ce_anchor}
\end{equation}
A flow level is sampled for each minibatch, and sequence-specific Dirichlet states provide the noisy inputs; padding is excluded. $\mathcal L_{\mathrm{CE}}$ denotes its training expectation. The distributed reduction averages over all valid nucleotides.

The formal PPO implementation sums log ratios over units, then clamps the joint log ratio to $[-20,20]$ before exponentiation for numerical stability; the PPO clipping range is a separate setting.

\paragraph{Empirical studies.}
The main configuration uses $G=8,H=8$.
Additional policy and training-group analyses are reported in Appendix~\ref{app:current_sensitivity}.
\subsection{Training data, parameter counts, and evaluation contracts}
\label{app:training_data_provenance}
\label{app:evaluation_contracts}
\label{app:reproducibility}
Supervised learning uses 10 million solver-generated sequence--structure pairs from the released RNA-Design-LM corpus. Post-training uses the released 2,790-target EternaWeb set with the supervised cross-entropy anchor retained. Sequence length ranges from 6 to 500 nucleotides. The original mixture uses all selected targets; the composition study groups them by the learnability-based categories in Appendix~\ref{app:figure_guide}.

The benchmark contract fixes target identities, target order, candidate order, and three evaluation seeds. 
Native RNA-IFlow uses 50 continuous-flow integration steps and terminal argmax decoding. 

Within each target--condition group, Pass@1 uses the first candidate and Pass@8 accepts any uMFE success among eight candidates. MFE@8 includes tied minima. Target probability, pair-set F1, and NED are scored per candidate and optimized independently within the group. No thermodynamic score guides sampling or reranking for the conditional generators. Diversity is the number of distinct valid returned RNA strings divided by eight. Empty failed returns do not count as unique sequences, but the eight-candidate denominator is retained.

Pair-set F1 is twice the number of shared predicted/target base pairs divided by the total number of pairs in the two sets. ViennaRNA's normalized ensemble defect is used directly, without a second division by length. The common final evaluator uses ViennaRNA 2.7.2 at 37 degrees Celsius with dangles 2 and unique multiloop decomposition. All benchmark sets are used solely for performance evaluation and do not participate in training, checkpoint choice, or hyperparameter tuning. Final benchmark metrics are reported on Eterna100-v2, Eterna100, and Rfam-27. The RNAsolo overlap analysis is given below. Cross-model supervised comparisons retain each model's trained weights and prescribed decoder.

\paragraph{Parameter accounting.}
Counts refer to deduplicated parameters in the evaluated model constructors and checkpoints. RNA-Design-LM uses a 13-token vocabulary. GoForth uses the pretrained-small configuration. DRAG includes its shared backbone and actor/critic heads, with 36,593 parameters in the actor-plus-backbone inference subset. Historical training masks for external models are unverified. Pure-search methods have no neural parameter count.
Model capacities and trainable parameter counts are summarized in Table~\ref{tab:parameter_counts}.
\begin{table}[htbp]
\centering\small
\setlength{\tabcolsep}{3pt}
\caption{This table reports model capacity. A dash denotes either no neural parameters for pure search or an unknown historical training mask.}
\label{tab:parameter_counts}
\begin{tabular}{lrS[table-format=3.2,round-precision=2]r}
\toprule
\multicolumn{1}{c}{Method} & \multicolumn{1}{c}{Total parameters} & \multicolumn{1}{c}{Millions} & \multicolumn{1}{c}{Trained parameters}\\
\midrule
RNA-Design-LM SL & 357921408 & 357.92 & {-}\\
RNA-Design-LM SL+RL & 357921408 & 357.92 & {-}\\
GoForth & \second{59392004} & \second{59.39} & {-}\\
DRAG & \best{37285} & \best{0.04} & {-}\\
RNAinverse-pf & {-} & {-} & {-}\\
RNA-IFlow & 87548613 & 87.55 & \second{86958021}\\
RNA-IFlow-RL & 87548613 & 87.55 & \best{14494917}\\
FastDesign & {-} & {-} & {-}\\
SAMFEO & {-} & {-} & {-}\\
NEMO & {-} & {-} & {-}\\
\bottomrule
\end{tabular}
\end{table}
\paragraph{Two-stage training and regularization.}
The finite policy is initialized from the supervised RNA-IFlow predictor. Each logical policy update advances four target structures with two fresh rollout rounds at the recorded group size. Earlier backbone layers and the reference remain frozen. The final two backbone blocks and output head are updated using terminal advantages, clipped PPO, structured-reference KL, and clean-nucleotide CE. The reward assigns weights 0.5 to target probability and 0.25 to each MFE and uMFE indicator. The clipping threshold is 0.2, with KL coefficient 0.01 and CE coefficient 0.1. Precise mathematical reductions are given above. The group-relative advantage vanishes at a singleton group, so no ordinary $G=1$ policy score is reported.

\subsection{Complete fixed-budget results and transfer}
\label{app:capacity_full_metrics}
\label{app:benchmark_scope}
All rows below use the corresponding frozen eight-return evaluation records and the common final scorer. Native-search comparators retain their internal search costs. Prob. denotes best target probability; F1 denotes best pair-set F1; Div. is valid-sequence diversity. Each best-of-eight metric is aggregated separately. Complete fixed-budget results are reported in Tables~\ref{tab:complete_eterna100v2}, \ref{tab:complete_eterna100}, \ref{tab:complete_rfam27}, and \ref{tab:complete_rnasolo764}.
\begin{table}[htbp]
\centering\small
\setlength{\tabcolsep}{3pt}
\caption{This table reports complete Eterna100-v2 results for eight returned candidates. Lower NED is better; all other metrics are higher-is-better.}
\label{tab:complete_eterna100v2}
\begin{tabular}{lS[table-format=1.4]S[table-format=1.4]S[table-format=1.4]S[table-format=1.4]S[table-format=1.4]S[table-format=1.4]S[table-format=1.4]S[table-format=1.4]}
\toprule
\multicolumn{1}{c}{Method} & \multicolumn{1}{c}{P@1} & \multicolumn{1}{c}{P@8} & \multicolumn{1}{c}{MFE@1} & \multicolumn{1}{c}{MFE@8} & \multicolumn{1}{c}{Prob.} & \multicolumn{1}{c}{F1} & \multicolumn{1}{c}{NED} & \multicolumn{1}{c}{Div.}\\
\midrule
RNA-Design-LM SL & 0.2633 & 0.4333 & 0.2700 & 0.4433 & 0.3364 & 0.7735 & 0.1694 & \best{0.9996}\\
RNA-Design-LM SL+RL & \second{0.5133} & \second{0.6367} & \second{0.5233} & \second{0.6567} & \second{0.5256} & \second{0.9324} & \second{0.0652} & 0.9783\\
GoForth & 0.3933 & 0.5700 & 0.4133 & 0.5833 & 0.3499 & 0.9236 & 0.0804 & 0.9862\\
DRAG & 0.3633 & 0.5133 & 0.3633 & 0.5200 & 0.2391 & 0.9062 & 0.1345 & 0.8029\\
RNAinverse-pf & 0.4367 & 0.5433 & 0.4533 & 0.5700 & 0.4344 & 0.6544 & 0.3625 & 0.6429\\
\textbf{RNA-IFlow (Ours)} & 0.3033 & 0.5200 & 0.3133 & 0.5367 & 0.4089 & 0.8731 & 0.1017 & \second{0.9979}\\
\textbf{RNA-IFlow-RL (Ours)} & \best{0.5400} & \best{0.6500} & \best{0.5433} & \best{0.6733} & \best{0.5398} & \best{0.9396} & \best{0.0534} & 0.9363\\
\bottomrule
\end{tabular}
\end{table}
\begin{table}[htbp]
\centering\small
\setlength{\tabcolsep}{3pt}
\caption{This table reports complete Eterna100 results for eight returned candidates. Lower NED is better; all other metrics are higher-is-better.}
\label{tab:complete_eterna100}
\begin{tabular}{lS[table-format=1.4]S[table-format=1.4]S[table-format=1.4]S[table-format=1.4]S[table-format=1.4]S[table-format=1.4]S[table-format=1.4]S[table-format=1.4]}
\toprule
\multicolumn{1}{c}{Method} & \multicolumn{1}{c}{P@1} & \multicolumn{1}{c}{P@8} & \multicolumn{1}{c}{MFE@1} & \multicolumn{1}{c}{MFE@8} & \multicolumn{1}{c}{Prob.} & \multicolumn{1}{c}{F1} & \multicolumn{1}{c}{NED} & \multicolumn{1}{c}{Div.}\\
\midrule
RNA-Design-LM SL & 0.2633 & 0.4267 & 0.2700 & 0.4333 & 0.3326 & 0.7669 & 0.1702 & \best{0.9996}\\
RNA-Design-LM SL+RL & \second{0.5033} & \second{0.6133} & \second{0.5167} & \second{0.6333} & \second{0.5165} & \second{0.9196} & \second{0.0691} & 0.9812\\
GoForth & 0.3933 & 0.5667 & 0.4133 & 0.5733 & 0.3435 & 0.9116 & 0.0848 & 0.9858\\
DRAG & 0.3633 & 0.5133 & 0.3633 & 0.5167 & 0.2376 & 0.8960 & 0.1373 & 0.8063\\
RNAinverse-pf & 0.4433 & 0.5100 & 0.4567 & 0.5267 & 0.4292 & 0.6442 & 0.3594 & 0.6400\\
\textbf{RNA-IFlow (Ours)} & 0.3000 & 0.4867 & 0.3100 & 0.5067 & 0.3988 & 0.8596 & 0.1099 & \second{0.9975}\\
\textbf{RNA-IFlow-RL (Ours)} & \best{0.5067} & \best{0.6167} & \best{0.5233} & \best{0.6467} & \best{0.5283} & \best{0.9247} & \best{0.0593} & 0.9333\\
\bottomrule
\end{tabular}
\end{table}
\begin{table}[htbp]
\centering\small
\setlength{\tabcolsep}{3pt}
\caption{This table reports complete Rfam-27 results for eight returned candidates. Lower NED is better; all other metrics are higher-is-better.}
\label{tab:complete_rfam27}
\begin{tabular}{lS[table-format=1.4]S[table-format=1.4]S[table-format=1.4]S[table-format=1.4]S[table-format=1.4]S[table-format=1.4]S[table-format=1.4]S[table-format=1.4]}
\toprule
\multicolumn{1}{c}{Method} & \multicolumn{1}{c}{P@1} & \multicolumn{1}{c}{P@8} & \multicolumn{1}{c}{MFE@1} & \multicolumn{1}{c}{MFE@8} & \multicolumn{1}{c}{Prob.} & \multicolumn{1}{c}{F1} & \multicolumn{1}{c}{NED} & \multicolumn{1}{c}{Div.}\\
\midrule
RNA-Design-LM SL & 0.5185 & 0.7531 & 0.5185 & 0.7531 & 0.6200 & 0.9814 & 0.0267 & \best{1.0000}\\
RNA-Design-LM SL+RL & \second{0.8148} & \second{0.8642} & \second{0.8148} & \second{0.8642} & \second{0.7765} & \best{0.9970} & \second{0.0058} & \best{1.0000}\\
GoForth & 0.6296 & 0.7284 & 0.6296 & 0.7407 & 0.4878 & 0.9930 & 0.0162 & \best{1.0000}\\
DRAG & 0.7160 & 0.8272 & 0.7407 & 0.8272 & 0.4218 & 0.9962 & 0.0468 & 0.5170\\
RNAinverse-pf & 0.3827 & 0.4568 & 0.3827 & 0.4568 & 0.4375 & 0.4932 & 0.5076 & 0.4398\\
\textbf{RNA-IFlow (Ours)} & 0.6543 & 0.8272 & 0.6667 & 0.8395 & 0.7158 & 0.9944 & 0.0104 & \best{1.0000}\\
\textbf{RNA-IFlow-RL (Ours)} & \best{0.8519} & \best{0.8765} & \best{0.8519} & \best{0.8765} & \best{0.7785} & \second{0.9969} & \best{0.0057} & \second{0.9907}\\
\bottomrule
\end{tabular}
\end{table}
\begin{table}[htbp]
\centering\small
\setlength{\tabcolsep}{3pt}
\caption{This table reports complete RNAsolo-764 results for eight returned candidates. Lower NED is better; all other metrics are higher-is-better.}
\label{tab:complete_rnasolo764}
\begin{tabular}{lS[table-format=1.4]S[table-format=1.4]S[table-format=1.4]S[table-format=1.4]S[table-format=1.4]S[table-format=1.4]S[table-format=1.4]S[table-format=1.4]}
\toprule
\multicolumn{1}{c}{Method} & \multicolumn{1}{c}{P@1} & \multicolumn{1}{c}{P@8} & \multicolumn{1}{c}{MFE@1} & \multicolumn{1}{c}{MFE@8} & \multicolumn{1}{c}{Prob.} & \multicolumn{1}{c}{F1} & \multicolumn{1}{c}{NED} & \multicolumn{1}{c}{Div.}\\
\midrule
RNA-Design-LM SL & 0.5951 & 0.7256 & 0.6069 & 0.7378 & 0.6418 & 0.9651 & 0.0347 & \best{0.9999}\\
RNA-Design-LM SL+RL & \best{0.7199} & \best{0.7531} & \best{0.7356} & \best{0.7657} & \best{0.7156} & \best{0.9808} & \best{0.0154} & 0.9774\\
GoForth & 0.6252 & 0.7216 & 0.6440 & 0.7347 & 0.5858 & 0.9756 & 0.0274 & 0.9848\\
\textbf{RNA-IFlow (Ours)} & 0.6165 & \second{0.7395} & 0.6339 & 0.7535 & 0.6890 & 0.9755 & 0.0203 & \second{0.9992}\\
\textbf{RNA-IFlow-RL (Ours)} & \second{0.7120} & 0.7312 & \second{0.7299} & \second{0.7544} & \second{0.7109} & \second{0.9783} & \second{0.0157} & 0.9186\\
\bottomrule
\end{tabular}
\end{table}
\label{tab:rnasolo_comparison}
RNAsolo separates first-candidate quality from sample coverage: post-training improves the base flow's P@1 and ensemble metrics, while the base model retains a larger distinct-sequence fraction and slightly higher P@8. RNA-Design-LM SL+RL has the highest P@8 on this collection. Nine exact target-structure overlaps occur in the supervised corpus. Excluding those targets without using outcomes leaves 755 targets and preserves the observed ordering; the exclusion is based on exact target-structure identity.

For the paired Eterna100-v2 comparison at $K=8$, the RNA-IFlow-RL minus RNA-Design-LM SL+RL success difference is 1.33 percentage points, with a target-bootstrap 95\% interval of $[-2.67,5.67]$. Targets are resampled with their evaluation conditions kept together. The interval includes zero.
The RNAsolo overlap sensitivity analysis is reported in Table~\ref{tab:rnasolo_overlap}.
\begin{table}[htbp]
\centering\small
\setlength{\tabcolsep}{3pt}
\caption{RNAsolo overlap sensitivity. The 755-target subset excludes the nine exact supervised target-structure overlaps without using model outcomes.}
\label{tab:rnasolo_overlap}
\begin{tabular}{lS[table-format=1.4]S[table-format=1.4]S[table-format=1.4]S[table-format=1.4]}
\toprule
\multicolumn{1}{c}{Method} & \multicolumn{1}{c}{Full P@8} & \multicolumn{1}{c}{755 P@8} & \multicolumn{1}{c}{Full NED} & \multicolumn{1}{c}{755 NED}\\
\midrule
RNA-Design-LM SL & 0.7256 & 0.7223 & 0.0347 & 0.0351\\
RNA-Design-LM SL+RL & \best{0.7531} & \best{0.7501} & \best{0.0154} & \best{0.0156}\\
GoForth & 0.7216 & 0.7183 & 0.0274 & 0.0274\\
RNA-IFlow & \second{0.7395} & \second{0.7364} & 0.0203 & 0.0205\\
RNA-IFlow-RL & 0.7312 & 0.7280 & \second{0.0157} & \second{0.0159}\\
\bottomrule
\end{tabular}
\end{table}

\begin{table}[htbp]
\centering\scriptsize
\setlength{\tabcolsep}{2.2pt}
\caption{This table reports overall performance as mean $\pm$ sample SD across three evaluation seeds. Bold and underlined means denote the best and second-best results, respectively.}
\label{tab:main_results_sd}
\begin{adjustbox}{width=\textwidth,center}
\begin{tabular}{lllcccccc}
\toprule
\multirow{2}{*}{Method} & \multirow{2}{*}{Type} & \multirow{2}{*}{Params (M)} & \multicolumn{2}{c}{Eterna100-v2} & \multicolumn{2}{c}{Eterna100} & \multicolumn{2}{c}{Rfam-27}\\
\cmidrule(lr){4-5}\cmidrule(lr){6-7}\cmidrule(lr){8-9}
&&& P@1 $\uparrow$ & P@8 $\uparrow$ & P@1 $\uparrow$ & P@8 $\uparrow$ & P@1 $\uparrow$ & P@8 $\uparrow$\\
\midrule
RNA-Design-LM SL & SL & 357.92 & 0.2633 $\pm$ 0.0231 & 0.4333 $\pm$ 0.0321 & 0.2633 $\pm$ 0.0231 & 0.4267 $\pm$ 0.0289 & 0.5185 $\pm$ 0.0370 & 0.7531 $\pm$ 0.0771\\
RNA-Design-LM SL+RL & SL+RL & 357.92 & \underline{0.5133} $\pm$ 0.0231 & \underline{0.6367} $\pm$ 0.0115 & \underline{0.5033} $\pm$ 0.0231 & \underline{0.6133} $\pm$ 0.0058 & \underline{0.8148} $\pm$ 0.0370 & \underline{0.8642} $\pm$ 0.0214\\
GoForth & SL & 59.39 & 0.3933 $\pm$ 0.0115 & 0.5700 $\pm$ 0.0000 & 0.3933 $\pm$ 0.0115 & 0.5667 $\pm$ 0.0058 & 0.6296 $\pm$ 0.0000 & 0.7284 $\pm$ 0.0214\\
\midrule
DRAG & Learned search & 0.04 & 0.3633 $\pm$ 0.0153 & 0.5133 $\pm$ 0.0058 & 0.3633 $\pm$ 0.0153 & 0.5133 $\pm$ 0.0058 & 0.7160 $\pm$ 0.0566 & 0.8272 $\pm$ 0.0214\\
RNAinverse-pf & Search & -- & 0.4367 $\pm$ 0.0058 & 0.5433 $\pm$ 0.0115 & 0.4433 $\pm$ 0.0208 & 0.5100 $\pm$ 0.0100 & 0.3827 $\pm$ 0.0566 & 0.4568 $\pm$ 0.0428\\
\midrule
\textbf{RNA-IFlow (Ours)} & SL & 87.55 & 0.3033 $\pm$ 0.0306 & 0.5200 $\pm$ 0.0000 & 0.3000 $\pm$ 0.0100 & 0.4867 $\pm$ 0.0416 & 0.6543 $\pm$ 0.0566 & 0.8272 $\pm$ 0.0214\\
\textbf{RNA-IFlow-RL (Ours)} & SL+RL & 87.55 & \textbf{0.5400} $\pm$ 0.0265 & \textbf{0.6500} $\pm$ 0.0000 & \textbf{0.5067} $\pm$ 0.0231 & \textbf{0.6167} $\pm$ 0.0058 & \textbf{0.8519} $\pm$ 0.0000 & \textbf{0.8765} $\pm$ 0.0214\\
\bottomrule
\end{tabular}
\end{adjustbox}
\end{table}

\subsection{Policy interface, horizon, and group-size analyses}
\label{app:current_sensitivity}
\label{app:negative_results}
The core policy comparison is reported in Table~\ref{tab:core_ablations}. Single- and three-seed H/G sensitivity, update-time profiling, native-flow step sensitivity, and post-training checkpoints are reported in Tables~\ref{tab:hg_sensitivity}, \ref{tab:hg_three_seed}, \ref{tab:g_profile}, \ref{tab:flow_steps}, and \ref{tab:training_updates}, respectively.

\begin{table}[htbp]
\centering\small
\setlength{\tabcolsep}{3pt}
\caption{This table summarizes the policy interface and post-training results on Eterna100-v2.}
\label{tab:core_ablations}
\begin{tabular}{lS[table-format=1.4]S[table-format=1.4]S[table-format=1.4]S[table-format=1.4]}
\toprule
\multicolumn{1}{c}{Generator} & \multicolumn{1}{c}{P@1} & \multicolumn{1}{c}{P@8} & \multicolumn{1}{c}{Prob.} & \multicolumn{1}{c}{NED}\\
\midrule
Native RNA-IFlow & \second{0.3033} & \second{0.5200} & \second{0.4089} & \second{0.1017}\\
RNA-IFlow-no-RL & 0.1233 & 0.2433 & 0.1822 & 0.2399\\
RNA-IFlow-RL & \best{0.5400} & \best{0.6500} & \best{0.5398} & \best{0.0534}\\
\bottomrule
\end{tabular}
\end{table}
A shared $H=8,G=8$ control is used for both axes.
\begin{table}[htbp]
\centering\small
\setlength{\tabcolsep}{3pt}
\caption{This table reports complete single-seed policy sensitivity; group size one is the zero-advantage boundary.}
\label{tab:hg_sensitivity}
\begin{tabular}{lrS[table-format=1.4]S[table-format=1.4]S[table-format=1.4]S[table-format=1.4]}
\toprule
\multicolumn{1}{c}{Axis} & \multicolumn{1}{c}{Value} & \multicolumn{1}{c}{P@1} & \multicolumn{1}{c}{P@8} & \multicolumn{1}{c}{Prob.} & \multicolumn{1}{c}{NED}\\
\midrule
H & 1 & 0.4533 & 0.6033 & 0.4821 & 0.0736\\
H & 2 & \second{0.5000} & \best{0.6567} & 0.5312 & \second{0.0590}\\
H & 4 & 0.4900 & 0.6367 & 0.5258 & 0.0598\\
H & 8 & \best{0.5167} & \second{0.6500} & \best{0.5424} & \best{0.0558}\\
H & 16 & \second{0.5000} & \second{0.6500} & \second{0.5317} & 0.0599\\
G & 2 & \second{0.5167} & 0.6500 & 0.5312 & 0.0587\\
G & 4 & 0.5100 & \second{0.6533} & 0.5339 & \best{0.0553}\\
G & 8 & \second{0.5167} & 0.6500 & \best{0.5424} & 0.0558\\
G & 16 & \best{0.5300} & \best{0.6600} & \second{0.5364} & \second{0.0557}\\
G & 1 & {N/A} & {N/A} & {N/A} & {N/A}\\
\bottomrule
\end{tabular}
\end{table}
\begin{table}[htbp]
\centering\small
\setlength{\tabcolsep}{3pt}
\caption{Values are reported as means and sample standard deviations across three repeated sensitivity runs.}
\label{tab:hg_three_seed}
\begin{tabular}{lcccc}
\toprule
\multicolumn{1}{c}{Configuration} & \multicolumn{1}{c}{P@1} & \multicolumn{1}{c}{P@8} & \multicolumn{1}{c}{Prob.} & \multicolumn{1}{c}{NED}\\
\midrule
$H=2,G=8$ & $0.5033\pm0.0058$ & $0.6511\pm0.0096$ & $0.5309\pm0.0063$ & $0.0593\pm0.0014$\\
$H=8,G=8$ & \second{$0.5122\pm0.0051$} & \best{$0.6578\pm0.0107$} & \best{$0.5391\pm0.0043$} & \best{$0.0554\pm0.0016$}\\
$H=8,G=4$ & \second{$0.5122\pm0.0102$} & \second{$0.6544\pm0.0051$} & $0.5351\pm0.0023$ & $0.0562\pm0.0008$\\
$H=8,G=16$ & \best{$0.5178\pm0.0158$} & \best{$0.6578\pm0.0019$} & \second{$0.5365\pm0.0016$} & \second{$0.0557\pm0.0013$}\\
\bottomrule
\end{tabular}
\end{table}
\begin{table}[htbp]
\centering\small
\setlength{\tabcolsep}{3pt}
\caption{This table reports observed group-size update costs on shared GPUs; minimum and maximum values summarize profiling variability.}
\label{tab:g_profile}
\begin{tabular}{rS[table-format=3.4]S[table-format=3.4]S[table-format=3.4]}
\toprule
\multicolumn{1}{c}{G} & \multicolumn{1}{c}{Median (s)} & \multicolumn{1}{c}{Minimum (s)} & \multicolumn{1}{c}{Maximum (s)}\\
\midrule
2 & \best{25.7918} & \second{10.5854} & \best{34.0818}\\
4 & \second{27.4301} & \best{10.4608} & \second{35.9115}\\
8 & 39.0918 & 11.9554 & 72.1897\\
16 & 56.9375 & 12.3717 & 302.8546\\
\bottomrule
\end{tabular}
\end{table}
\begin{table}[htbp]
\centering\small
\setlength{\tabcolsep}{3pt}
\caption{This table reports native-flow integration-step sensitivity on 100 Eterna100-v2 targets.}
\label{tab:flow_steps}
\begin{tabular}{rS[table-format=1.4]S[table-format=1.4]S[table-format=1.4]S[table-format=1.4]}
\toprule
\multicolumn{1}{c}{Steps} & \multicolumn{1}{c}{P@1} & \multicolumn{1}{c}{P@8} & \multicolumn{1}{c}{MFE@8} & \multicolumn{1}{c}{Prob.}\\
\midrule
4 & 0.2800 & 0.5033 & 0.5033 & 0.3991\\
8 & 0.2900 & \second{0.5133} & \best{0.5367} & 0.4041\\
16 & 0.2967 & \second{0.5133} & \best{0.5367} & 0.4056\\
32 & \second{0.3000} & \second{0.5133} & \second{0.5333} & \best{0.4095}\\
50 & \best{0.3033} & \best{0.5200} & \best{0.5367} & \second{0.4089}\\
\bottomrule
\end{tabular}
\end{table}
\begin{table}[htbp]
\centering\small
\setlength{\tabcolsep}{3pt}
\caption{This table reports checkpoint performance during post-training continuation.}
\label{tab:training_updates}
\begin{tabular}{rS[table-format=1.4]S[table-format=1.4]S[table-format=1.4]S[table-format=1.4]}
\toprule
\multicolumn{1}{c}{Updates} & \multicolumn{1}{c}{P@1} & \multicolumn{1}{c}{P@8} & \multicolumn{1}{c}{MFE@8} & \multicolumn{1}{c}{Prob.}\\
\midrule
1744 & 0.5233 & \best{0.6567} & \best{0.6867} & \best{0.5446}\\
2093 & \second{0.5333} & \second{0.6533} & 0.6767 & 0.5364\\
2442 & \best{0.5400} & 0.6500 & 0.6733 & 0.5398\\
2790 & 0.5300 & 0.6500 & \second{0.6800} & \second{0.5437}\\
\bottomrule
\end{tabular}
\end{table}
\subsection{Data composition, fixed exposure, and regularization}
\label{app:composition_details}
The composition study matches target count, length distribution, initialization, and 352 logical updates. The coverage study fixes 4,144 target visits over 1,036 logical updates while changing the number of distinct targets. Both are single-training-seed studies. The original mixture and its display categories are described in Appendix~\ref{app:figure_guide}. Coverage, regularization, and rollout-freshness controls are summarized in Tables~\ref{tab:d1_coverage_revisit}, \ref{tab:regularization}, and \ref{tab:refresh_corrected}.
\begin{table}[htbp]
\centering\small
\setlength{\tabcolsep}{3pt}
\caption{This table compares matched-size training-data compositions.}
\label{tab:composition_corrected}
\begin{tabular}{lS[table-format=1.4]S[table-format=1.4]S[table-format=1.4]S[table-format=1.4]S[table-format=1.4]}
\toprule
\multicolumn{1}{c}{Setting} & \multicolumn{1}{c}{P@1} & \multicolumn{1}{c}{P@8} & \multicolumn{1}{c}{Prob.} & \multicolumn{1}{c}{NED} & \multicolumn{1}{c}{Div.}\\
\midrule
Original mixture & \best{0.5133} & \best{0.6533} & \best{0.5363} & \best{0.0555} & \best{0.9513}\\
Core-only & 0.4867 & \second{0.6333} & 0.5203 & 0.0616 & 0.9279\\
Hard-enriched & \second{0.4900} & \second{0.6333} & \second{0.5248} & \second{0.0610} & \second{0.9379}\\
\bottomrule
\end{tabular}
\end{table}
\begin{table}[htbp]
\centering\small
\setlength{\tabcolsep}{3pt}
\caption{This table reports coverage under a fixed number of target visits.}
\label{tab:d1_coverage_revisit}
\begin{tabular}{lS[table-format=1.4]S[table-format=1.4]S[table-format=1.4]S[table-format=1.4]S[table-format=1.4]}
\toprule
\multicolumn{1}{c}{Setting} & \multicolumn{1}{c}{P@1} & \multicolumn{1}{c}{P@8} & \multicolumn{1}{c}{Prob.} & \multicolumn{1}{c}{NED} & \multicolumn{1}{c}{Div.}\\
\midrule
1395 unique tasks & 0.5000 & \best{0.6500} & \second{0.5361} & \second{0.0590} & 0.9071\\
2790 unique tasks & \best{0.5133} & 0.6267 & 0.5269 & 0.0596 & \second{0.9175}\\
4143 unique tasks & \second{0.5033} & \second{0.6467} & \best{0.5367} & \best{0.0565} & \best{0.9292}\\
\bottomrule
\end{tabular}
\end{table}
\begin{table}[htbp]
\centering\small
\setlength{\tabcolsep}{3pt}
\caption{This table reports one-axis regularization sensitivity with all other settings fixed.}
\label{tab:regularization}
\begin{tabular}{lS[table-format=1.4]S[table-format=1.4]S[table-format=1.4]S[table-format=1.4]S[table-format=1.4]}
\toprule
\multicolumn{1}{c}{Setting} & \multicolumn{1}{c}{P@1} & \multicolumn{1}{c}{P@8} & \multicolumn{1}{c}{Prob.} & \multicolumn{1}{c}{NED} & \multicolumn{1}{c}{Div.}\\
\midrule
Lower CE weight & 0.5133 & 0.6433 & 0.5332 & 0.0576 & \second{0.9487}\\
Higher CE weight & \second{0.5167} & \second{0.6467} & \second{0.5361} & 0.0582 & \best{0.9517}\\
Lower KL weight & \second{0.5167} & \best{0.6533} & \best{0.5408} & \best{0.0546} & 0.9413\\
Higher KL weight & \best{0.5233} & 0.6433 & 0.5313 & \second{0.0561} & 0.9400\\
\bottomrule
\end{tabular}
\end{table}
Rollout refresh aligns newly sampled trajectories with the latest policy. The snapshot variants collect distinct candidate batches from a fixed behavior snapshot within a logical update. The two- and three-round comparisons retain their corresponding rollout and optimizer budgets. Extra fresh rounds perform extra work, so accuracy differences across round counts describe the combined schedule.
\begin{table}[htbp]
\centering\small
\setlength{\tabcolsep}{3pt}
\caption{This table reports rollout-freshness and reuse controls.}
\label{tab:refresh_corrected}
\begin{tabular}{lS[table-format=1.4]S[table-format=1.4]S[table-format=1.4]S[table-format=1.4]}
\toprule
\multicolumn{1}{c}{Setting} & \multicolumn{1}{c}{P@1} & \multicolumn{1}{c}{P@8} & \multicolumn{1}{c}{Prob.} & \multicolumn{1}{c}{NED}\\
\midrule
One fresh round & 0.5133 & 0.6633 & 0.5425 & 0.0551\\
Two fresh rounds & \second{0.5167} & \second{0.6500} & \best{0.5424} & \second{0.0558}\\
Fixed behavior snapshot, two rounds & \best{0.5233} & \best{0.6533} & \second{0.5396} & \best{0.0557}\\
Three fresh rounds & \second{0.5100} & \second{0.6467} & \second{0.5396} & \second{0.0558}\\
Fixed behavior snapshot, three rounds & \best{0.5233} & \best{0.6600} & \best{0.5435} & \best{0.0554}\\
\bottomrule
\end{tabular}
\end{table}

\subsection{Timing and nested candidate budgets}
\label{app:budget_runtime}
Neural timing keeps the evaluated model resident on the same A100. All 100 Eterna100-v2 targets and their three fixed sampling conditions are measured twice with rotated method order. The generation and post-generation scoring intervals are measured separately. Candidate fingerprints were checked against the frozen eight-candidate ledger. Detailed return and scoring times are reported in Table~\ref{tab:runtime_full}. A search algorithm's time includes its internal folding and search, so the native rows below describe end-to-end candidate return rather than an equivalent neural forward pass.
\begin{table}[htbp]
\centering\small
\setlength{\tabcolsep}{3pt}
\caption{This table reports time to return eight candidates and time after common scoring. Neural entries use matched resident-model medians, while search rows retain their native timing scope.}
\label{tab:runtime_full}
\begin{tabular}{lS[table-format=3.4]S[table-format=3.4]r}
\toprule
\multicolumn{1}{c}{Method} & \multicolumn{1}{c}{Return time (s)} & \multicolumn{1}{c}{With scoring (s)} & \multicolumn{1}{c}{Timed groups}\\
\midrule
RNA-Design-LM SL & 4.4085 & 4.7356 & 600\\
RNA-Design-LM SL+RL & 4.4085 & 4.6438 & 600\\
GoForth & 1.3518 & 1.6502 & 600\\
DRAG & 14.8637 & 15.0512 & 300\\
RNAinverse-pf & 133.5565 & 133.7930 & 300\\
RNA-IFlow & \second{0.8845} & \second{1.1684} & 600\\
RNA-IFlow-RL & \best{0.7317} & \best{0.9186} & 600\\
\bottomrule
\end{tabular}
\end{table}
The nested-budget experiment extends the frozen candidate prefix rather than resampling independently for each budget. A point is admitted only when every target and condition has been scored. Historical missing generation times remain unreported. Nested-prefix results for RNA-Design-LM SL, RNA-Design-LM SL+RL, GoForth, RNA-IFlow, and RNA-IFlow-RL are reported in Tables~\ref{tab:nested_RNADesignLMSL}, \ref{tab:nested_RNADesignLMSLRL}, \ref{tab:nested_GoForth}, \ref{tab:nested_RNAIFlow}, and \ref{tab:nested_RNAIFlowRL}. Identity-checked prefix-replay time plus new-block time is a separate measurement and is not used as a same-GPU cross-model speed comparison.
\begin{table}[htbp]
\centering\small
\setlength{\tabcolsep}{3pt}
\caption{This table reports RNA-Design-LM SL nested-prefix results from the same ordered candidate set, with each metric aggregated separately.}
\label{tab:nested_RNADesignLMSL}
\begin{tabular}{rS[table-format=1.4]S[table-format=1.4]S[table-format=1.4]S[table-format=1.4]S[table-format=1.4]S[table-format=1.4]}
\toprule
\multicolumn{1}{c}{K} & \multicolumn{1}{c}{Pass@K} & \multicolumn{1}{c}{MFE@K} & \multicolumn{1}{c}{Prob.} & \multicolumn{1}{c}{NED} & \multicolumn{1}{c}{Pair-F1} & \multicolumn{1}{c}{Diversity}\\
\midrule
1 & 0.2633 & 0.2700 & 0.2136 & 0.2762 & 0.5899 & \best{1.0000}\\
2 & 0.3433 & 0.3500 & 0.2618 & 0.2276 & 0.6662 & \best{1.0000}\\
4 & 0.4033 & 0.4100 & 0.3024 & 0.1946 & 0.7313 & \best{1.0000}\\
8 & 0.4333 & 0.4433 & 0.3364 & 0.1694 & 0.7735 & \second{0.9996}\\
16 & 0.4800 & 0.4967 & 0.3719 & 0.1497 & 0.8187 & 0.9992\\
32 & 0.5167 & 0.5467 & 0.4064 & 0.1319 & 0.8539 & 0.9985\\
64 & 0.5667 & 0.5933 & 0.4304 & 0.1198 & 0.8758 & 0.9971\\
128 & 0.5900 & 0.6167 & 0.4502 & 0.1105 & 0.8936 & 0.9957\\
256 & \second{0.6067} & \second{0.6333} & \second{0.4663} & \second{0.1017} & \second{0.9101} & 0.9937\\
512 & \best{0.6400} & \best{0.6633} & \best{0.4748} & \best{0.0950} & \best{0.9173} & 0.9916\\
\bottomrule
\end{tabular}
\end{table}
\begin{table}[htbp]
\centering\small
\setlength{\tabcolsep}{3pt}
\caption{This table reports RNA-Design-LM SL+RL nested-prefix results from the same ordered candidate set, with each metric aggregated separately.}
\label{tab:nested_RNADesignLMSLRL}
\begin{tabular}{rS[table-format=1.4]S[table-format=1.4]S[table-format=1.4]S[table-format=1.4]S[table-format=1.4]S[table-format=1.4]}
\toprule
\multicolumn{1}{c}{K} & \multicolumn{1}{c}{Pass@K} & \multicolumn{1}{c}{MFE@K} & \multicolumn{1}{c}{Prob.} & \multicolumn{1}{c}{NED} & \multicolumn{1}{c}{Pair-F1} & \multicolumn{1}{c}{Diversity}\\
\midrule
1 & 0.5133 & 0.5233 & 0.4238 & 0.1099 & 0.8319 & \best{1.0000}\\
2 & 0.5833 & 0.5900 & 0.4756 & 0.0855 & 0.8847 & \second{0.9950}\\
4 & 0.6200 & 0.6333 & 0.5036 & 0.0743 & 0.9145 & 0.9883\\
8 & 0.6367 & 0.6567 & 0.5256 & 0.0652 & 0.9324 & 0.9783\\
16 & 0.6500 & 0.6800 & 0.5375 & 0.0582 & 0.9493 & 0.9683\\
32 & 0.6600 & 0.6967 & 0.5501 & 0.0530 & 0.9543 & 0.9581\\
64 & 0.6733 & 0.7100 & 0.5597 & 0.0503 & 0.9609 & 0.9443\\
128 & 0.7000 & 0.7367 & 0.5671 & 0.0479 & 0.9634 & 0.9308\\
256 & 0.7167 & 0.7567 & 0.5748 & 0.0454 & 0.9684 & 0.9168\\
512 & \second{0.7367} & \second{0.7700} & \second{0.5824} & \second{0.0437} & \second{0.9712} & 0.9016\\
1024 & \best{0.7600} & \best{0.7967} & \best{0.5875} & \best{0.0419} & \best{0.9748} & 0.8864\\
\bottomrule
\end{tabular}
\end{table}
\begin{table}[htbp]
\centering\small
\setlength{\tabcolsep}{3pt}
\caption{This table reports GoForth nested-prefix results from the same ordered candidate set, with each metric aggregated separately.}
\label{tab:nested_GoForth}
\begin{tabular}{rS[table-format=1.4]S[table-format=1.4]S[table-format=1.4]S[table-format=1.4]S[table-format=1.4]S[table-format=1.4]}
\toprule
\multicolumn{1}{c}{K} & \multicolumn{1}{c}{Pass@K} & \multicolumn{1}{c}{MFE@K} & \multicolumn{1}{c}{Prob.} & \multicolumn{1}{c}{NED} & \multicolumn{1}{c}{Pair-F1} & \multicolumn{1}{c}{Diversity}\\
\midrule
1 & 0.3933 & 0.4133 & 0.2302 & 0.1357 & 0.8096 & \best{1.0000}\\
2 & 0.4633 & 0.4800 & 0.2714 & 0.1101 & 0.8667 & \second{0.9933}\\
4 & 0.5433 & 0.5533 & 0.3276 & 0.0895 & 0.9075 & 0.9900\\
8 & 0.5700 & 0.5833 & 0.3499 & 0.0804 & 0.9236 & 0.9862\\
16 & 0.5933 & 0.6200 & 0.3731 & 0.0744 & 0.9360 & 0.9831\\
32 & 0.6100 & 0.6300 & 0.3894 & 0.0693 & 0.9435 & 0.9793\\
64 & 0.6267 & 0.6400 & 0.4062 & 0.0656 & 0.9483 & 0.9736\\
128 & 0.6433 & 0.6567 & 0.4170 & 0.0625 & 0.9534 & 0.9668\\
256 & 0.6600 & 0.6800 & 0.4317 & 0.0597 & 0.9583 & 0.9595\\
512 & \second{0.6700} & \second{0.6900} & \second{0.4398} & \second{0.0573} & \second{0.9614} & 0.9523\\
1024 & \best{0.6833} & \best{0.7067} & \best{0.4494} & \best{0.0554} & \best{0.9644} & 0.9433\\
\bottomrule
\end{tabular}
\end{table}
\begin{table}[htbp]
\centering\small
\setlength{\tabcolsep}{3pt}
\caption{This table reports RNA-IFlow nested-prefix results from the same ordered candidate set, with each metric aggregated separately.}
\label{tab:nested_RNAIFlow}
\begin{tabular}{rS[table-format=1.4]S[table-format=1.4]S[table-format=1.4]S[table-format=1.4]S[table-format=1.4]S[table-format=1.4]}
\toprule
\multicolumn{1}{c}{K} & \multicolumn{1}{c}{Pass@K} & \multicolumn{1}{c}{MFE@K} & \multicolumn{1}{c}{Prob.} & \multicolumn{1}{c}{NED} & \multicolumn{1}{c}{Pair-F1} & \multicolumn{1}{c}{Diversity}\\
\midrule
1 & 0.3033 & 0.3133 & 0.2534 & 0.1880 & 0.6872 & \best{1.0000}\\
2 & 0.3733 & 0.3900 & 0.3076 & 0.1477 & 0.7766 & \best{1.0000}\\
4 & 0.4600 & 0.4800 & 0.3680 & 0.1224 & 0.8334 & 0.9975\\
8 & 0.5200 & 0.5367 & 0.4089 & 0.1017 & 0.8731 & \second{0.9979}\\
16 & 0.5700 & 0.5967 & 0.4410 & 0.0894 & 0.9020 & 0.9960\\
32 & 0.5933 & 0.6167 & 0.4650 & 0.0818 & 0.9175 & 0.9931\\
64 & 0.6100 & 0.6367 & 0.4844 & 0.0756 & 0.9290 & 0.9904\\
128 & 0.6467 & 0.6767 & 0.5037 & 0.0699 & 0.9371 & 0.9875\\
256 & 0.6600 & 0.6900 & 0.5165 & 0.0657 & 0.9422 & 0.9852\\
512 & \second{0.6700} & \second{0.7100} & \second{0.5273} & \second{0.0611} & \second{0.9491} & 0.9821\\
1024 & \best{0.6800} & \best{0.7200} & \best{0.5359} & \best{0.0584} & \best{0.9541} & 0.9793\\
\bottomrule
\end{tabular}
\end{table}
\begin{table}[htbp]
\centering\small
\setlength{\tabcolsep}{3pt}
\caption{This table reports RNA-IFlow-RL nested-prefix results from the same ordered candidate set, with each metric aggregated separately.}
\label{tab:nested_RNAIFlowRL}
\begin{tabular}{rS[table-format=1.4]S[table-format=1.4]S[table-format=1.4]S[table-format=1.4]S[table-format=1.4]S[table-format=1.4]}
\toprule
\multicolumn{1}{c}{K} & \multicolumn{1}{c}{Pass@K} & \multicolumn{1}{c}{MFE@K} & \multicolumn{1}{c}{Prob.} & \multicolumn{1}{c}{NED} & \multicolumn{1}{c}{Pair-F1} & \multicolumn{1}{c}{Diversity}\\
\midrule
1 & 0.5400 & 0.5433 & 0.4483 & 0.0906 & 0.8527 & \best{1.0000}\\
2 & 0.5867 & 0.5900 & 0.4854 & 0.0748 & 0.8906 & \second{0.9850}\\
4 & 0.6233 & 0.6300 & 0.5127 & 0.0633 & 0.9190 & 0.9683\\
8 & 0.6500 & 0.6733 & 0.5398 & 0.0534 & 0.9396 & 0.9363\\
16 & 0.6633 & 0.6933 & 0.5517 & 0.0499 & 0.9510 & 0.9071\\
32 & 0.6833 & 0.7133 & 0.5626 & 0.0462 & 0.9597 & 0.8772\\
64 & 0.6967 & 0.7267 & 0.5725 & 0.0430 & 0.9644 & 0.8485\\
128 & 0.6967 & 0.7267 & 0.5804 & 0.0406 & 0.9673 & 0.8197\\
256 & 0.7133 & 0.7433 & 0.5849 & 0.0386 & 0.9719 & 0.7908\\
512 & \second{0.7167} & \second{0.7467} & \second{0.5891} & \second{0.0375} & \second{0.9739} & 0.7612\\
1024 & \best{0.7233} & \best{0.7567} & \best{0.5925} & \best{0.0363} & \best{0.9768} & 0.7337\\
\bottomrule
\end{tabular}
\end{table}
\subsection{Native-search quality and computation}
\label{app:native_search_60s}
FastDesign, SAMFEO, and NEMO are evaluated locally on Eterna100-v2 using 100 targets and three seeds, with a maximum native search time of 60 seconds per target and seed. All returned sequences are rescored with the common ViennaRNA 2.7.2 implementation. These methods may evaluate many internal candidates and return different numbers of designs.  The eight-return main table and this fixed-time native-search comparison have separate budgets.

FastDesign uses the authors' fast configuration without optional rival-structure search. SAMFEO uses the documented development-branch implementation. NEMO retains its native energy settings, and all final metrics use the common scorer. Every planned target--seed group remains in the denominator. Native-search performance and completion statistics are reported in Tables~\ref{tab:native_search_complete} and \ref{tab:native_search_coverage}. A group with no returned sequence has zero success, zero target probability, unit NED, and zero pair-set F1 by the declared failure convention.  A timeout may still have returned candidates and is distinguished from an empty return.
\begin{table}[htbp]
\centering\small
\setlength{\tabcolsep}{3pt}
\caption{This table reports native-search performance on Eterna100-v2 with at most 60 seconds per target and seed. Time is the observed mean search time, and all 300 groups contribute to the metrics.}
\label{tab:native_search_complete}
\begin{tabular}{lS[table-format=1.4]S[table-format=1.4]S[table-format=1.4]S[table-format=1.4]S[table-format=1.4]S[table-format=2.4]}
\toprule
\multicolumn{1}{c}{Method} & \multicolumn{1}{c}{uMFE} & \multicolumn{1}{c}{MFE} & \multicolumn{1}{c}{Prob.} & \multicolumn{1}{c}{NED} & \multicolumn{1}{c}{Pair-F1} & \multicolumn{1}{c}{Run time (s)}\\
\midrule
FastDesign & \second{0.4900} & \second{0.5267} & \second{0.3827} & 0.3589 & \second{0.6623} & \second{33.5269}\\
SAMFEO & \best{0.6667} & \best{0.6933} & \best{0.5283} & \best{0.0579} & \best{0.9474} & 43.3045\\
NEMO & 0.2933 & 0.3167 & 0.1720 & \second{0.2928} & 0.6530 & \best{15.9501}\\
\bottomrule
\end{tabular}
\end{table}
\begin{table}[htbp]
\centering\small
\setlength{\tabcolsep}{3pt}
\caption{This table reports search completion and returned-candidate counts under the same native 60-second protocol. Returned counts describe the final candidate sets.}
\label{tab:native_search_coverage}
\begin{tabular}{lrrrS[table-format=2.4]}
\toprule
\multicolumn{1}{c}{Method} & \multicolumn{1}{c}{Groups} & \multicolumn{1}{c}{Timeouts} & \multicolumn{1}{c}{Empty returns} & \multicolumn{1}{c}{Mean returned}\\
\midrule
FastDesign & 300 & \second{93} & 93 & 3.4267\\
SAMFEO & 300 & 172 & \best{0} & 9.9200\\
NEMO & 300 & \best{68} & \second{48} & 0.8400\\
\bottomrule
\end{tabular}
\end{table}
SAMFEO has a slightly higher uMFE success rate than the eight-candidate RNA-IFlow-RL protocol, while RNA-IFlow-RL has higher target probability and lower NED. The comparison therefore describes a quality--compute trade-off. Figure~\ref{fig:search_frontier} displays the runtime distributions and arithmetic means. Neural generation and native search retain their respective hardware and internal scoring costs. Internal oracle counts were not instrumented for the native runs.
\subsection{Published and large-budget solution coverage}
\label{app:published_provenance}
\label{app:large_budget}
Large-budget results complement the local 60-second search comparison. For RNA-IFlow-RL, the maximum budget is 10,000 candidates per target across sampling streams, with global stopping after uMFE success. Counts measure targets solved by any stream, with the stated candidate allowance serving as an upper limit. Large-budget coverage is summarized in Tables~\ref{tab:large_budget_eterna100}, \ref{tab:large_budget_transfer}, and \ref{tab:published_rnasolo_search}. Published search counts preserve each source's folding engine, repeats, and budget.
\begin{table}[htbp]
\centering\small
\setlength{\tabcolsep}{3pt}
\caption{This table reports Eterna100 large-budget coverage as solved targets out of 100. NEMO counts follow the SAMFEO authors' reproduction.}
\label{tab:large_budget_eterna100}
\begin{tabular}{lrrp{.43\textwidth}}
\toprule
\multicolumn{1}{c}{Method} & \multicolumn{1}{c}{MFE} & \multicolumn{1}{c}{uMFE} & \multicolumn{1}{c}{Budget / source}\\
\midrule
RNA-IFlow-RL & 73 & 68 & At most 10,000 candidates; global stop\\
RNA-Design-LM SL+RL & 74 & 72 & Published 10,000-sample evaluation~\citep{gautam2026rnadesignlm}\\
FastDesign & \best{80} & \best{78} & Published motif/cube search~\citep{zhou2026fastdesign}\\
SAMFEO & 77 & 74 & Five-run union, probability objective~\citep{zhou2023samfeo}\\
NEMO & \second{79} & \second{77} & Five-run union reported by~\citet{zhou2023samfeo}\\
\bottomrule
\end{tabular}
\end{table}
\begin{table}[htbp]
\centering\small
\setlength{\tabcolsep}{3pt}
\caption{This table reports additional large-budget generator coverage under each source's original stopping and aggregation protocol.}
\label{tab:large_budget_transfer}
\begin{tabular}{llrrp{.25\textwidth}}
\toprule
\multicolumn{1}{c}{Benchmark} & \multicolumn{1}{c}{Method} & \multicolumn{1}{c}{MFE} & \multicolumn{1}{c}{uMFE} & \multicolumn{1}{c}{Protocol}\\
\midrule
Eterna100-v2 & RNA-IFlow-RL & \best{81} & \second{76} & Maximum 10,000 candidates\\
Eterna100-v2 & RNA-Design-LM SL+RL & \best{81} & \best{79} & Published 10,000 samples~\citep{gautam2026rnadesignlm}\\
Rfam-27 & RNA-IFlow-RL & \best{24} & \best{24} & Maximum 10,000 candidates\\
Rfam-27 & RNA-Design-LM SL+RL & \best{24} & \best{24} & Published 10,000 samples~\citep{gautam2026rnadesignlm}\\
RNAsolo-764 & RNA-IFlow-RL & 591 & 585 & Maximum 10,000 candidates\\
\bottomrule
\end{tabular}
\end{table}
For FastDesign, the reported Eterna100 configuration uses 5,000 motif iterations, 2,500 root iterations, and cube-pruning size 90. Its published mean time is 284.1000 seconds on the authors' CPU system. SAMFEO's probability-objective experiment uses an ensemble size of 10, a temperature of 1, and 5,000 search iterations, with five independent runs. Its reported union of 74 uMFE solutions differs from its mean per-run count. NEMO's 77 uMFE solutions in the same comparison are also a five-run union~\citep{zhou2023samfeo,zhou2026fastdesign}. SamplingDesign reports 76 uMFE solutions on Eterna100 under its continuous-search protocol~\citep{tang2026samplingdesign}. The reported coverage depends on the search budget and stopping rule.
\begin{table}[htbp]
\centering\small
\setlength{\tabcolsep}{3pt}
\caption{This table reports published RNAsolo-764 search coverage from the FastDesign comparison using the source paper's native search protocol.}
\label{tab:published_rnasolo_search}
\begin{tabular}{lrrl}
\toprule
\multicolumn{1}{c}{Method} & \multicolumn{1}{c}{MFE solved} & \multicolumn{1}{c}{uMFE solved} & \multicolumn{1}{c}{Source}\\
\midrule
FastDesign & \best{611} & \best{606} & \citet{zhou2026fastdesign}, Table 2\\
SAMFEO & 608 & 602 & \citet{zhou2026fastdesign}, Table 2\\
NEMO & \second{609} & \second{605} & \citet{zhou2026fastdesign}, Table 2\\
\bottomrule
\end{tabular}
\end{table}

\subsection{Reference, selection rule, and structural interpretation}
\label{app:five_method_case}
\label{app:structure_examples}
\label{app:hard_cases}
The main-text example is Eterna puzzle 8935338, puzzle 82 (Anemone), with a 214-nucleotide target. The reference is the author/player-designed sequence in the original ``Sample Solution (V2/Vienna2)'' field and its paired ``Secondary Structure V2'' record. The sequence and structure are not cropped. The original record is available in the Eterna benchmarking repository\footnote{\url{https://github.com/eternagame/eterna100-benchmarking}.}. It is a designed reference rather than a natural RNA sequence.

All five methods use the same target and condition with eight saved candidates. Within each method, the displayed candidate maximizes target probability, breaking ties by candidate index. The target is a post-hoc discordant case where RNA-IFlow-RL achieves uMFE success and the four displayed comparison methods have no target MFE hit within their corresponding groups. Population-level comparisons are reported in Tables~\ref{tab:main_results} and~\ref{tab:thermodynamic_efficiency}. Each panel shows its own predicted MFE topology. Different sequence letters relative to the designed reference are marked separately from correct, alternative, or missing structure edges.

\begin{table}[htbp]
\centering\small
\setlength{\tabcolsep}{3pt}
\caption{This table reports displayed-candidate thermodynamics and group-level hit counts for Anemone. Reference metrics are evaluated independently and do not form a generated $K=8$ group.}
\label{tab:case_metrics}
\begin{tabular}{lS[table-format=1.4]S[table-format=1.4]cccc}
\toprule
\multicolumn{1}{c}{Sequence} & \multicolumn{1}{c}{Prob.} & \multicolumn{1}{c}{NED} & \multicolumn{1}{c}{MFE} & \multicolumn{1}{c}{uMFE} & \multicolumn{1}{c}{Group MFE} & \multicolumn{1}{c}{Group uMFE}\\
\midrule
Reference sequence & 0.1803 & 0.0367 & yes & yes & -- & --\\
RNA-IFlow-RL & \best{0.5178} & \best{0.0128} & yes & yes & \best{2/8} & \best{2/8}\\
RNA-IFlow & \second{0.1257} & 0.0773 & no & no & \second{0/8} & \second{0/8}\\
RNA-Design-LM SL+RL & 0.0769 & \second{0.0297} & no & no & \second{0/8} & \second{0/8}\\
GoForth & 0.0019 & 0.1477 & no & no & \second{0/8} & \second{0/8}\\
DRAG & 0.0007 & 0.0447 & no & no & \second{0/8} & \second{0/8}\\
\bottomrule
\end{tabular}
\end{table}
\paragraph{Exact sequences and structures.}
The numbered blocks below are contiguous segments of each sequence, followed by its displayed structure. The reference structure is the requested target. The complete strings and source identities are retained in the accompanying figure data.

\paragraph{Target (214 nt).}\mbox{}\par
\begingroup\small\ttfamily
\noindent 001--054 AAAGGUUCGGAAACGAGCCGAUGGGAAACCAUCGGUCAGGAAACUGGCCGAGGG\\
\noindent 055--108 GAAACCCUCGCACAGGAAACUGUGCGUAACGAAAGUUACCGUGCGAAAGCACGC\\
\noindent 109--162 GCCCCGAAAGGGGUGACUGGGGAAACCCGGGAUUAGGAAACUAAUCGGGCGGAA\\
\noindent 163--214 ACGCCCGUUUUGGAAACAAGACGUUGGGAAACCAACGCGUCCGAAAGGACGC\\
\normalfont\small Displayed structure:\par\ttfamily
\noindent ...((((((....))))))(((((....)))))((((((....))))))(((((\\
\noindent ....)))))((((((....))))))(((((....)))))(((((....)))))(\\
\noindent (((((....)))))).(((((....)))))((((((....))))))(((((...\\
\noindent .)))))((((((....))))))(((((....)))))((((((....))))))\\
\endgroup

\paragraph{RNA-IFlow-RL (214 nt).}\mbox{}\par
\begingroup\small\ttfamily
\noindent 001--054 AAAGCUGGCAAAAGCCAGCGCACCGAAAGGUGCGCCGACGAAAGUCGGCGCCGG\\
\noindent 055--108 GAAACCGGCGCCCGCGAAAGCGGGCGGCCCGAAAGGGCCGGGCCGAAAGGCCCG\\
\noindent 109--162 GGGCCGAAAGGCCCCAGGGGCGAAAGCCCCGCUGUCGAAAGACAGCGACUCGAA\\
\noindent 163--214 AGAGUCGCGGGCGAAGGCCCGCGGUCCGAAAGGACCGCCGUCGAAGGACGGC\\
\normalfont\small Displayed structure:\par\ttfamily
\noindent ...((((((....))))))(((((....)))))((((((....))))))(((((\\
\noindent ....)))))((((((....))))))(((((....)))))(((((....)))))(\\
\noindent (((((....)))))).(((((....)))))((((((....))))))(((((...\\
\noindent .)))))((((((....))))))(((((....)))))((((((....))))))\\
\endgroup

\paragraph{RNA-IFlow (214 nt).}\mbox{}\par
\begingroup\small\ttfamily
\noindent 001--054 AAAGCCGGCACUAGCCGGCGCGGCAUAAGCCGCGUCCCCGAAUGGGGACGGGGC\\
\noindent 055--108 AAGGGCCCCGGGCGCAAAAGCGCCCGGGCCAAAAGGCCCGCCCCAAAAGGGGCG\\
\noindent 109--162 ACACCAAAAGGUGUCACCGCCAGGUGGCGGGGCGGCAAAAGCCGCCGGCCCAAA\\
\noindent 163--214 AGGGCCCAGCGCUCACGCGCUGGAUCCGAAAGGGUCGCGACCAUAAGGUCGC\\
\normalfont\small Displayed structure:\par\ttfamily
\noindent ...((((((....))))))(((((....(((((((((((....))))))(((((\\
\noindent ....)))))((((((....))))))(((((....)))))(((((....)))))(\\
\noindent (((((....)))))).(((((....))))).)))))....))))).(((((...\\
\noindent .)))))((((((....))))))(((((....)))))((((((....))))))\\
\endgroup

\paragraph{RNA-Design-LM SL+RL (214 nt).}\mbox{}\par
\begingroup\small\ttfamily
\noindent 001--054 AAAGCCCUCACAAGAGGGCCAUCCAAAAGGAUGACGAGGGUGACCUCGUGGGGC\\
\noindent 055--108 AAAAGCCCCGAUCCCAAAAGGGAUUCAGCCAAAAGGCUGGACCCGAAAGGGUCG\\
\noindent 109--162 CCAUGGAGACAUGGCACCGAGGAAACUCGGGCCCGCGAACGCGGGCGCCACACA\\
\noindent 163--214 CGUGGCCCAUGGGAAACCGUGGCAACCCAAGGGUUGGGAUGGGAAACCAUCC\\
\normalfont\small Displayed structure:\par\ttfamily
\noindent ..(((((((....))))))(((((....)))))((((((....))))))(((((\\
\noindent ....)))))((((((....))))))(((((....)))))(((((....)))))(\\
\noindent (((((....)))))).(((((....)))))((((((....))))))(((((...\\
\noindent .)))))((((((....))))))..(((....)))).((((((....))))))\\
\endgroup

\paragraph{GoForth (214 nt).}\mbox{}\par
\begingroup\small\ttfamily
\noindent 001--054 AUUGGGUGCAUUAGCACCCGGGUGAAAUCACCCGUUGGGUAAUCCCGACGCUGU\\
\noindent 055--108 AAUUACAGCGUUGGUAACAACCAACUUGGCACACGCCAAGCGGGUAAUCCCGCG\\
\noindent 109--162 GGUGUAACUACACCCAGCUGGAUAUCCAGCGGGGUCAACAGACCCCGGGCUAUA\\
\noindent 163--214 AAGCCCGUGUGCAGAAGCACACGUCGGAACUCCGACGCUUGGGAUGCCAAGC\\
\normalfont\small Displayed structure:\par\ttfamily
\noindent ...((((((....))))))(((((....)))))((((((.....((((((((((\\
\noindent ....)))))((((((....))))))(((((....)))))(((((....)))))(\\
\noindent (((((....)))))).(((((....)))))((((((....))))))(((((...\\
\noindent .)))))((((((....)))))))))))...))))))((((((....))))))\\
\endgroup

\paragraph{DRAG (214 nt).}\mbox{}\par
\begingroup\small\ttfamily
\noindent 001--054 ACAGGGAGGGAUACCUCCCGGGCGAAAACGCCCGGGGGCCAAAGCCCCCGGGAG\\
\noindent 055--108 AAAACUCCCGGUGGGAAAACCCACCGUGGGAAAACCCACUGGGGCAACCCCCAG\\
\noindent 109--162 CGGGGAAAACCCCGCAGUGGCAAAAGCCACGGGGUGUUAACACCCCGGCGGUAA\\
\noindent 163--214 UCCGCCGGGGGGAAAACCCCCCGGGGGGAAACCCCCGGCGGUAAAGACCGCC\\
\normalfont\small Displayed structure:\par\ttfamily
\noindent .(.((((((....))))))(((((....)))))((((((....))))))(((((\\
\noindent ....)))))((((((....))))))(((((....)))))(((((....)))))(\\
\noindent (((((....)))))).(((((....)))))((((((....))))))(((((...\\
\noindent .)))))((((((....))))))(((((....))))))(((((....))))).\\
\endgroup
\paragraph{Failure modes across the evaluated targets.}
A candidate can retain many target base pairs yet fail uMFE because an alternative complete structure has lower or equal energy. It can also attain uMFE while allocating appreciable equilibrium probability to near-optimal competitors. These cases motivate reporting both success and ensemble quality. The native-flow and finite-policy controls, paired uncertainty, and RNAsolo results above provide complementary failure analyses at the population level. 

\subsection{Finite-policy sampling and post-training procedure}
\label{app:algorithm}
The following procedure summarizes supervised training, post-training, and inference.
\begin{enumerate}[leftmargin=*,itemsep=3pt]
\item Fit the clean-nucleotide predictor to supervised sequence--structure pairs along the Dirichlet path. Initialize the refinement model with these learned parameters.
\item For a target, initialize each unpaired unit uniformly over four nucleotides and each paired unit uniformly over the six legal pair states.
\item At each policy step, map the current discrete state to the Dirichlet conditional mean. Obtain temperature-scaled nucleotide probabilities, construct the SPD unit distributions, and sample the keep-or-resample transition independently across units conditional on the complete current sequence. Record the behavior probability of each executed transition.
\item After all policy steps, score the completed sequences with target probability, MFE success, and uMFE success. Standardize rewards within each trajectory group and assign the resulting terminal advantage to every transition of that trajectory.
\item Compute each step's joint current-to-behavior ratio across all units. Apply the clipped surrogate with a sum over policy steps and an average over trajectories, together with structured-reference KL and the supervised CE anchor.
\item Refresh rollout data for subsequent policy updates. During inference, execute the finite policy directly for each requested candidate, without reward evaluation or score-based reranking.
\end{enumerate}
\subsection{Additional recovered and residual examples}
\label{app:additional_cases}
\begin{figure}[htbp]\centering
\includegraphics[width=\textwidth]{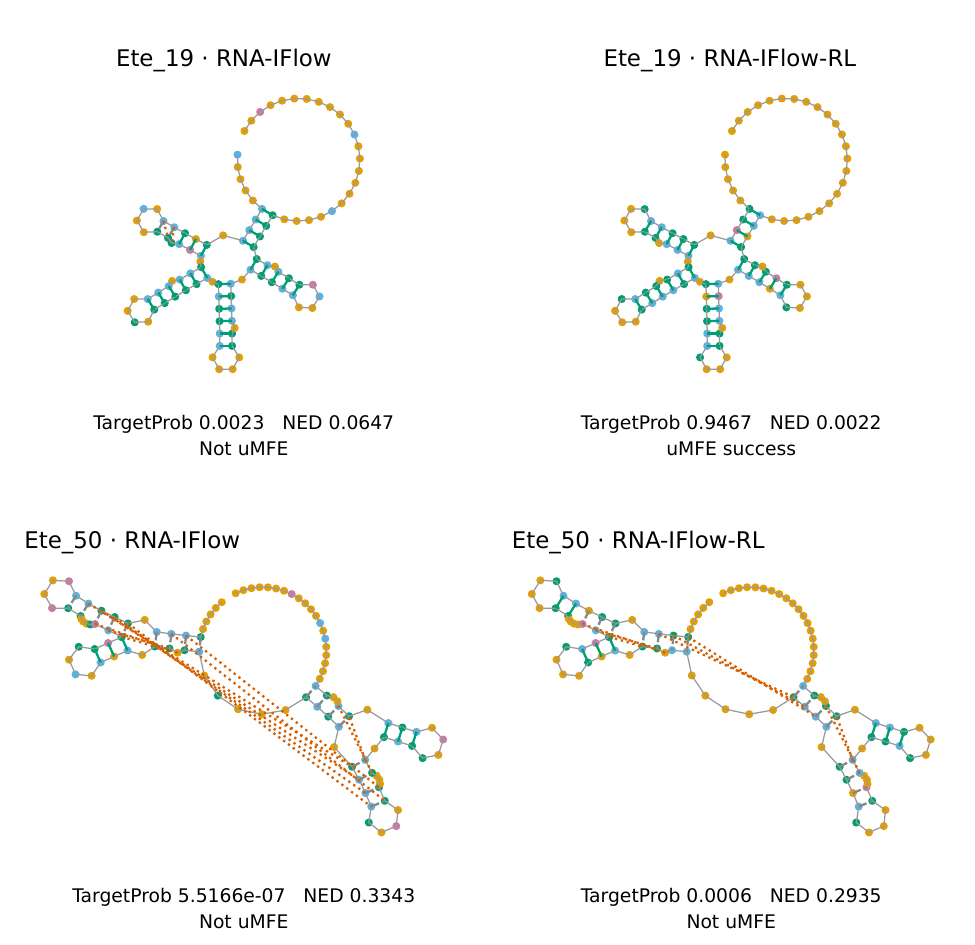}
\caption{This figure shows additional first-candidate designs, with a recovered target above and a residual uMFE failure below. Green, orange, and dashed gray bonds denote recovered, alternative, and missing pairs, while nucleotide colors denote A/C/G/U in orange/blue/green/purple.}
\label{fig:additional_cases}
\end{figure}

Two further examples use the first candidate at fixed seed 1009. Within the strata ``native flow fails and post-training succeeds'' and ``both fail'', the target of median length is selected, breaking ties by target identifier. The resulting 104- and 108-nucleotide targets were chosen without maximizing improvement. These are saved full-pipeline outputs, so the native-to-refined comparison changes both weights and generation dynamics. Figure~\ref{fig:additional_cases} shows these additional designs, and Table~\ref{tab:additional_case_metrics} reports their thermodynamic metrics.
\begin{table}[htbp]
\centering\small
\setlength{\tabcolsep}{3pt}
\caption{This table reports additional fixed-first-candidate thermodynamics; small nonzero probabilities are retained in scientific notation.}
\label{tab:additional_case_metrics}
\begin{tabular}{lrlS[table-format=1.4]S[table-format=1.4]c}
\toprule
\multicolumn{1}{c}{Target} & \multicolumn{1}{c}{nt} & \multicolumn{1}{c}{Model} & \multicolumn{1}{c}{Prob.} & \multicolumn{1}{c}{NED} & \multicolumn{1}{c}{uMFE}\\
\midrule
Ete\_19 & 104 & RNA-IFlow & \second{0.0023} & \second{0.0647} & no\\
Ete\_19 & 104 & RNA-IFlow-RL & \best{0.9467} & \best{0.0022} & yes\\
Ete\_50 & 108 & RNA-IFlow & {$5.5166\times10^{-7}$} & \second{0.3343} & no\\
Ete\_50 & 108 & RNA-IFlow-RL & 0.0006 & \best{0.2935} & no\\
\bottomrule
\end{tabular}
\end{table}
\subsection{Historical training-signal profiles}
\label{app:training_signal}
\begin{figure}[htbp]\centering
\includegraphics[width=\textwidth]{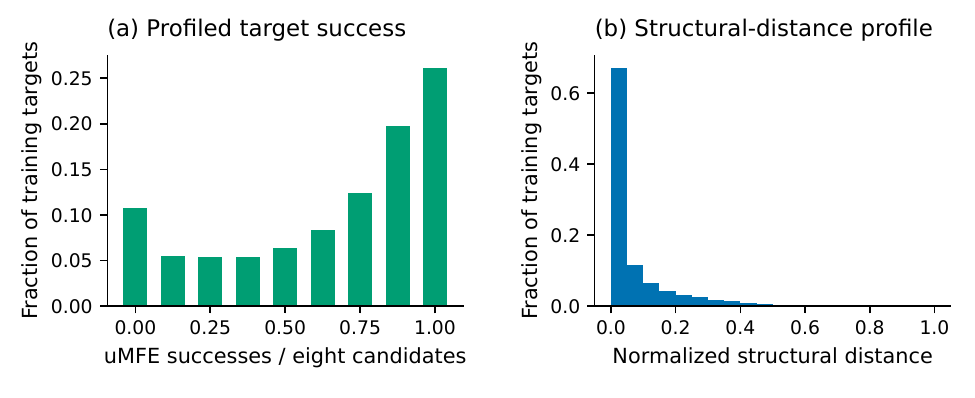}
\caption{This figure shows frozen-supervised-flow profiles over 12,613 training-pool targets, each with eight valid candidates. Panels (a,b) report target fractions by empirical uMFE success and mean normalized structural distance, respectively.}
\label{fig:training_signal}
\end{figure}

The archived profiling experiment evaluates eight frozen-supervised-flow candidates for each of 12,617 training-pool targets. Its all-or-nothing (AoN) quantity is the fraction of candidates meeting the uMFE criterion, not a mean target probability. Its normalized structural distance (NSD) is the source evaluator's structure-distance function divided by target length, not a coefficient of variation. The visual summary retains the 12,613 targets for which all eight candidate records are valid. Four targets contain a total of 13 invalid records and are excluded from this distribution plot. This profiling collection was not itself used to select the final 2,790-target training mixture. Figure~\ref{fig:training_signal} summarizes the corresponding empirical training-signal profiles.

\subsection{Recorded training progress and complementary outcomes}
\label{app:recorded_training_process}
Gray points show per-update observations and the blue curve shows non-overlapping 32-update means.

\begin{figure}[htbp]
\centering
\includegraphics[width=.72\textwidth]{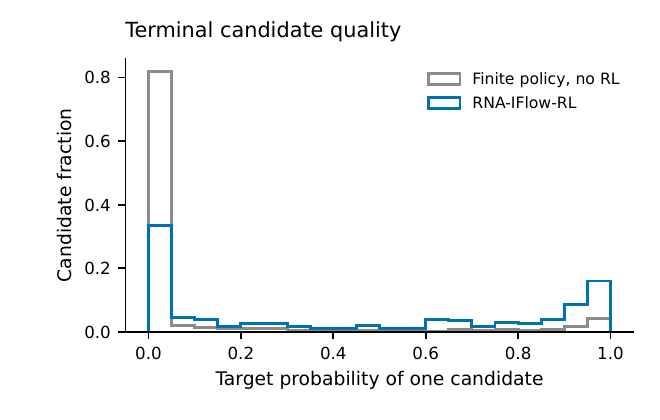}
\caption{This figure compares candidate-level target-probability histograms before and after RL. Gray and blue outlines denote the finite policy without RL and RNA-IFlow-RL, while the vertical axis reports the candidate fraction in each probability bin.}
\label{fig:terminal_distribution_supp}
\end{figure}

Rewards are measured on the changing training-target stream. The complementary histogram in Figure~\ref{fig:terminal_distribution_supp} displays the candidate-level target-probability frequencies underlying this distributional comparison. Schedule and reference-anchor controls are reported in Appendix~\ref{app:current_sensitivity}.

\subsection{Training-data length composition}
\label{app:length_composition}
\begin{figure}[htbp]
\centering
\includegraphics[width=.72\textwidth]{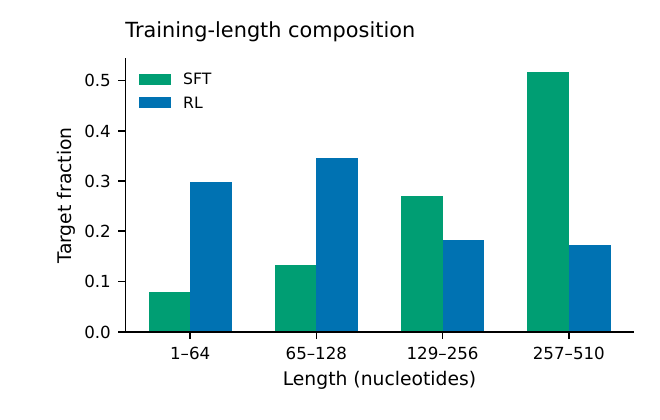}
\caption{This figure shows training-set length distributions. Bars report sample fractions in each length interval, with green and blue denoting supervised and RL training data, respectively.}
\label{fig:training_lengths_supp}
\end{figure}

The supervised and RL training-set length distributions are shown in Figure~\ref{fig:training_lengths_supp}. The matched-composition controls hold target counts and length-bin counts fixed; coverage controls instead hold total target visits fixed.

\subsection{Case-display window and full-length sequence recovery}
\label{app:recovery_window}
The structural case retains the original 214-nucleotide reference and all five generated candidates. Figure~\ref{fig:structure_refinement} displays the same central 50 positions, 83--132, for every sequence. The structures, candidate selection, target probability, NED, and MFE/uMFE status all use complete sequences. The display window is fixed across methods.

For the author-designed reference $x^{\mathrm{ref}}$, sequence recovery is the fraction of identical bases at aligned positions,
\begin{equation}
\operatorname{Recovery}(x,x^{\mathrm{ref}})
=\frac{1}{L}\sum_{j=1}^{L}\mathbf 1[x_j=x_j^{\mathrm{ref}}].
\label{eq:case_recovery}
\end{equation}
This measure characterizes resemblance to the player-designed reference. Folding success is evaluated separately. Structurally valid alternative sequences can differ substantially from the reference. Full-length recovery values are summarized in Table~\ref{tab:case_recovery}.
\begin{table}[htbp]
\centering\small
\caption{This table reports full-length recovery for the displayed Anemone candidates using the same 214-nucleotide reference throughout.}
\label{tab:case_recovery}
\begin{tabular}{lcr}
\toprule
\multicolumn{1}{c}{Sequence} & \multicolumn{1}{c}{Matching positions} & \multicolumn{1}{c}{Recovery (\%)}\\
\midrule
Author/player reference & 214/214 & 100.00\\
RNA-Design-LM SL+RL & 87/214 & 40.65\\
GoForth & 88/214 & 41.12\\
DRAG & \second{101/214} & \second{47.20}\\
RNA-IFlow & 97/214 & 45.33\\
RNA-IFlow-RL & \best{114/214} & \best{53.27}\\
\bottomrule
\end{tabular}
\end{table}

\subsection{Related work in context}
\label{app:related_work}
RNA inverse folding has developed along three directions: target-specific search, learned optimization, and conditional generation. Search methods exploit structural constraints and thermodynamic objectives, as in NEMO, SAMFEO, SamplingDesign, and FastDesign~\citep{portela2018nemo,zhou2023samfeo,tang2026samplingdesign,zhou2026fastdesign}. LEARNA and DRAG learn reusable construction or mutation policies, while RNA-Design-LM and GoForth generate sequences conditioned on structural specifications~\citep{runge2019learna,li2025drag,gautam2026rnadesignlm,lindsey2026goforth}.

Flow Matching learns probability transport, with Dirichlet and discrete formulations extending it to categorical sequences~\citep{lipman2023flow,stark2024dirichlet,gat2024dfm}. RNAFlow and RiboFlow apply flow-based models to RNA sequence--structure co-design~\citep{nori2024rnaflow,ma2025riboflow}. Reward-based methods such as DRAKES and flow-policy optimization further adapt generators to downstream objectives~\citep{wang2025drakes,mcallister2026fpo,su2026domino}. RNA-IFlow-RL brings these perspectives together under a fixed secondary-structure condition: global sequence variation is refined through a finite policy that coordinates paired actions and learns from terminal folding feedback.

\subsection{Discussion and future directions}
\label{app:discussion}
Our framework separates structural compatibility from thermodynamic preference. Coordinated sequence updates explore pairing-compatible candidates, while terminal feedback favors complete sequences whose folding ensembles support the target. The finite policy connects these two levels of design. The coverage and native-search analyses also show the importance of assessing success together with diversity and computational cost. Extending the feedback to alternative energy models and experimentally measured properties is a natural direction for testing whether the learned preferences transfer to biological settings.

\end{document}